\documentclass[twocolumn]{autart}    

\usepackage{graphicx}          
\usepackage[dvips]{epsfig}    
\usepackage{mathrsfs}
\usepackage{amsfonts}
\usepackage[fleqn]{amsmath}
\usepackage{amssymb}

\usepackage{float}

\usepackage[dvips]{color}
\usepackage{subfigure}
\usepackage{caption}

\def\dref#1{(\ref{#1})}

\newtheorem{remark}{Remark}

\begin{document}

\begin{frontmatter}

\title{Discrete Communication and Control Updating in Event-Triggered Consensus\thanksref{footnoteinfo}} 

\thanks[footnoteinfo]{This work was supported in part by the National Natural Science Foundation of China under grants 61473005 and 11332001. Corresponding author Z.~Li.}

\author[TJ,ZX]{Bin Cheng}\ead{bincheng@tongji.edu.cn},    
\author[BIT,YDRA]{Yuezu Lv}\ead{yzlv@bit.edu.cn},
\author[PKU]{Zhongkui Li}\ead{zhongkli@pku.edu.cn},               
\author[PKU]{Zhisheng Duan}\ead{duanzs@pku.edu.cn},                                                                

\address[TJ]{Department of Control Science and Engineering, College of Electronics and Information Engineering, Tongji University, Shanghai 201804, China}
\address[ZX]{Shanghai Research Institute for Intelligent Autonomous Systems, Shanghai 200120, China}
\address[BIT]{Advanced Research Institute of Multidisciplinary Sciences, Beijing Institute of Technology, Beijing 100081, China}
\address[YDRA]{Yangtze Delta Region Academy of Beijing Institute of Technology, Jiaxing 314000, China}
\address[PKU]{State Key Laboratory for Turbulence and Complex Systems, Department of Mechanics and Engineering Science, College of Engineering, Peking University, Beijing 100871, China} 

%
%


\begin{abstract}                          
This paper studies the consensus control problem faced with three essential demands, namely, discrete control updating for each agent, discrete-time communications among neighboring agents, and the fully distributed fashion of the controller implementation without requiring any global information of the whole network topology.
Noting that the existing related results only meeting one or two demands at most are essentially not applicable, in this paper we establish a novel framework to solve the problem of fully distributed consensus with discrete communication and control.
The first key point in this framework is the design of controllers that are only updated at discrete event instants and do not depend on global information by introducing time-varying gains inspired by the adaptive control technique.
Another key point is the invention of novel dynamic triggering functions that are independent of relative information among neighboring agents.
Under the established framework, we propose fully distributed state-feedback event-triggered protocols for undirected graphs and also further study the more complexed cases of output-feedback control and directed graphs.
Finally, numerical examples are provided to verify the effectiveness of the proposed event-triggered protocols.
\end{abstract}

\end{frontmatter}

\section{Introduction}\label{s1}
Information interactions among neighboring agents and control updating of each agent are not only necessary but essential to achieve cooperative control multi-agent systems.
Only from this point, a high frequency of interaction and control updating is conductive to increase the speed of convergence and obtain a good control performance.
However, every coin has two sides.
From the perspective of security, a high-frequency interaction often increases the risk of information leakage, which may make the whole systems paralyzed.
From the view of controller, a high-frequency updating may increase the fatigue of actuators and thus reduce their usable lives.
Actually, in engineering scenarios too high-frequency interactions are impractical,
since on-board agents often have limited energy and communication bandwidths.
Also notice that in multi-agent systems each agent only associates with its neighboring agents, so it is expected to design fully distributed protocols without requiring any global information of the network.

This paper aims at studying the problem of fully distributed consensus with discrete communication and control (FDCDCC).
We are going to propose fully distributed event-triggered protocols that only need discrete communications among neighboring agents, are discretely updated, and do not require any global information.
Limited by the challenges imposed by the above three factors, this is still an open and difficult research topic.
To the best of our knowledge, until now various scientists have made contributions to just one or two requirements of this problem; see \cite{DVDimarogonas2012distributed,Garcia2013decentralised,Heemels2013TAC,Meng2013event,Seyboth2013event,NaHuang2020IS,PYu2018TAC,Nowzari2019Auto} and the references therein.
In this paper, we intend to systematically analyze and thoroughly work out this challenging problem.

\subsection{Literature Review}
In the past decade, various researchers have paid their attentions to reducing the frequency of communication by applying the event-triggered strategy to multi-agent systems.
Researchers presented several kinds of event-based protocols in \cite{Garcia2013decentralised,Seyboth2013event,HZhang2014observer,Zhu2014event,DYang2016Decentralized,Liu2017distributed,thcheng2017event} to solve the consensus problem of multi-agent systems with linear models, including single integrators, double integratorss, and general linear system.
In \cite{XYi2019TAC,Girard2015TAC,Dolk2017TAC}, the authors designed dynamic event-triggered protocols to solve the distributed consensus problem.
In \cite{Cheng2016Event,Yu2016leader}, distributed event-triggered protocols were proposed to solve the leader-follower consensus in the presence of a dynamic leader.
References on event-triggered control from other perspectives have also been reported, such as disturbance rejection \cite{LXing2017event}, finite-time consensus \cite{HYu2020TAC}, output consensus of heterogeneous networks \cite{WHu2017output}. Synchronization of multi-agent systems was realized by using event-triggered and self-triggered broadcasts in \cite{Alm2017TAC}.

There were also some works considering how to reduce the frequency of control updating, like \cite{NaHuang2020IS}.
However, continuous communication could not be avoided in \cite{NaHuang2020IS} since the triggering functions were designed based on continuous information among neighbors.
In \cite{PYu2018TAC}, the authors used event triggering to reduce communication and controller updating for multi-agent systems.
However, the protocols proposed in \cite{PYu2018TAC} were only applicable to systems connected by undirected graph.

It is not difficult to find that in the above existing works, the proposed event-based protocols rely on some global value(s) associated with the whole network topology.
Generally, those global values are not easy to compute or even not available to each local agent, indicating that the protocols of these papers are not applicable to complex and large-scale networks.
Even if we only consider how to reduce communication frequency, it is nontrivial to give proper event-triggered protocols to achieve expected collective behavior in a fully distributed fashion.
To avoid requiring global information, the works \cite{BCheng2018fully,BCheng2019tac} introduced adaptive coupling gains into both controllers and event-triggered conditions.
In \cite{YangyangQian2019TAC}, the authors studied output consensus of heterogeneous linear multi-agent systems with adaptive event-triggered control.

Note that the protocols given in \cite{BCheng2018fully,BCheng2019tac,YangyangQian2019TAC} were designed for undirected topology and may not be applied in a general directed topology of which the Laplacian matrix is asymmetric.
In \cite{XianweiLi2020}, the authors proposed adaptive event-triggered algorithms for directed network topologies.
However, the controllers of \cite{XianweiLi2020} needed to be updated continuously because of the time-varying coupling gains introduced into the controllers.
Moreover, the triggering functions designed in \cite{XianweiLi2020} required continuous relative information among neighboring agents which did not coincide with the original idea of using event-triggered control to reduce communication frequency.

\subsection{Main Contribution}
This paper aims to solve the FDCDCC problem by proposing fully distributed event-triggered protocols that do not need continuous communication or updating.
The co-existence of the three constraints, i.e., fully distributed fashion, discrete communication, and discrete control updating, imposes several significant challenges. Specifically, these three coupled constraints result in structured nonlinear networks with asynchronous and aperiodic control and communication signals. This renders that the existing methods of designing control laws or triggering functions in e.g., \cite{BCheng2018fully,BCheng2019tac,YangyangQian2019TAC,XianweiLi2020} cannot be either directly applied or easily extended to simultaneously handle these three constraints.

In this paper, a novel event triggering framework is proposed to tackle the FDCDCC problem.
There are at least two key points in this framework.
The first one is the design of controllers that are only updated at discrete event instants and do not depend on any global information by introducing time-varying gains inspired by the adaptive control technique.
The other is the invention of novel dynamic triggering functions that are independent of relative information among neighboring agents.
Such triggering functions are decoupled with the network topology, which plays a key role in solving the FDCDCC problem.
Under the established framework, fully distributed event-triggered protocols are first presented to solve the FDCDCC problem under undirected graphs.
On this basis, we further solve the problem under general directed graphs with asymmetric Laplacian matrices which significantly increase the difficulty of the algorithm design and the corresponding theoretical analysis.
Extensions to the output-feedback case are also considered in this paper.

Different from \cite{NaHuang2020IS,PYu2018TAC} in which the event-triggered protocols either require continuous communication to check triggering conditions or only apply to undirected graphs, the proposed protocols given in this paper do not depend on continuous relative information among neighboring agents and can be used in general directed graphs.
Also notice that this paper is essentially different from the existing related papers \cite{BCheng2018fully,YangyangQian2019TAC}, which study fully distributed consensus control under undirected papers.
In the structures of \cite{BCheng2018fully,YangyangQian2019TAC}, it is hard to solve the FDCDCC problem constrained by discrete control updating and directed graphs.
Thus, the presented protocols in the current paper are of more general significance than the ones in \cite{NaHuang2020IS,PYu2018TAC,BCheng2018fully,YangyangQian2019TAC}.

It is to be pointed out that in \cite{BCheng2020CDC} we propose adaptive event-triggered protocols based on relative state measurements rather communications.
At costs, each agent needs to transmit the input information to its out-neighbors at event instants and compute an extra variable.
However, this method consumes extra computation cost and relies on accurate system dynamics.
These drawbacks essentially exist in the protocols based on relative state measurements, such as the ones of \cite{BCheng2020CDC,XianweiLi2020,XianweiLi2020Automatica}.
Such limitations can be circumvented by using the distributed adaptive dynamic event-triggered protocols proposed in the current paper.

\subsection{Organization of the Paper}
The rest of this paper is organized as follows.
We first formulate the problem in Section \ref{s2}.
Then, we study the undirected graph case and propose distributed adaptive event-triggered protocols in Section \ref{s3}.
Next, the case of general directed graphs is investigated in Section \ref{s4}.
In Section \ref{s5}, we further consider the case where state information is not available.
In Section \ref{s6}, we introduce the simulation results and make some discussions.
Finally, we conclude this paper in Section \ref{s7}.

\subsection{Notation}
$I_p$ represents the $p$-dimensional identity matrix and $\textbf{1}_p$ denotes a $p$-dimensional column vector with elements being $1$.
Symbol $\text{diag}(x_1,\cdots,x_n)$ represents a diagonal matrix with diagonal elements being $x_i$.
$\sigma_{\max}(X)$ represents the maximum singular values of the matrix $X$. For a symmetric matrix $Z$, $\lambda_{\max}(Z)$ and $\lambda_{\min}(Z)$ denote the maximum and minimal eigenvalues of $Z$, respectively.
$A\otimes B$ is the Kronecker product of $A$ and $B$.

\section{Problem Formulation}\label{s2}
Consider a group of $N$ agents with dynamics described by
\begin{equation}\label{model1}
\begin{aligned}
&\dot{x}_{i}=Ax_{i}+Bu_{i},\\
&y_i=Cx_i,~i=1,\cdots,N,
\end{aligned}
\end{equation}
where $x_i\in\mathbf{R}^n$, $y_i\in\mathbf{R}^m$, and $u_i\in\mathbf{R}^{p}$ represent the state, measured output, and control input of the $i$-th agent, respectively, and $A$, $B$, $C$ are constant known matrices with compatible dimensions. Furthermore, the triple $(A,B,C)$ is stabilizable and detectable.

The communication topology among the $N$ agents is represented by a directed graph $\mathcal {G}=(\mathcal {V}, \mathcal{E})$, where $\mathcal {V}=\{1,\cdots,N\}$ is the node set and $\mathcal {E}\subseteq\mathcal {V}\times\mathcal
{V}$ is the edge set, in which an edge is represented by an
ordered pair of distinct nodes. If $(i,j)\in\mathcal {E}$, node $i$ is called a neighbor of node $j$, and the latter is called an out-neighbor of the former.
A directed path from node ${i_1}$ to node ${i_l}$ is a sequence of adjacent edges of the form $({i_k}, {i_{k+1}})$, $k=1,\cdots,l-1$.
A directed graph is strongly connected if there is a directed path from every node to every other node.
A graph is said to be undirected, if $(i,j)\in\mathcal E$ as long as $(j,i)\in\mathcal E$.
An undirected graph is connected if there exists a path between every pair of distinct nodes, otherwise is disconnected.
For the graph $\mathcal G$, its adjacency matrix, denoted by $\mathcal A=[a_{ij}]\in \mathbf R^{N\times N}$, is defined such that $a_{ii}=0$, $a_{ij}=1$ if $(j,i)\in \mathcal E$ and $a_{ij}=0$ otherwise.
The Laplacian matrix $\mathcal L=[l_{ij}]\in \mathbf R^{(N\times N}$ associated with graph $\mathcal G$ is defined as $l_{ii}=\sum_{j=1}^N{a_{ij}}$ and $l_{ij}=-a_{ij}$, $i\neq j$.

In this paper, we aim at studying the problem of FDCDCC.
Three main constrains are involved in this problem, namely, limited updates of controllers, limited bandwidths of communications, and unavailable global information associated with the whole network.
The above three constrains are very common in numerous practical occasions, while the co-presence of them renders the problem quite challenging.
To the best knowledge of  the authors, no existing protocols can be used for this new and open problem.

In this paper, we are going to systematically study the problem of FDCDCC from a few different viewpoints, such as undirected or directed graph, state-based or output-based controller, leaderless or leader-follower consensus.
For each case, we shall design proper distributed protocols to realize consensus in the sense that $\lim_{t\rightarrow \infty}\|x_i(t)- x_j(t)\|=0$, $\forall\,i,j=1,\cdots,N$ and ensure that there does not exist the Zeno behavior \cite{DYang2016Decentralized}.

Let $\xi=[\xi_1^T,\cdots,\xi_{N}^T]^T$ where $\xi_i=\sum_{j=1}^Na_{ij}(x_i-x_j)$.
Then we have $\xi=(\mathcal{L}\otimes I)x$, where $x=[x_1^T,\cdots,x_{N}^T]^T$.
We refer to $\xi$ as the consensus error. It is clear that consensus is achieved if and only if $\xi$ asymptotically converges to zero.

%
%


\section{FDCDCC for Undirected Graphs}\label{s3}
In this section, we consider the problem of FDCDCC formulated in Section \ref{s2} under undirected graphs.




Due to the constraint of discrete communication and control updating, we cannot directly use the consensus error $\xi_i$ to design the control input $u_i$ as in the existing works like \cite{YuezuLv2016Automatica,ZLi2014cooperative}.
Instead, only the state information at event-triggering instant can be utilized to generate the controller. For convenience, denote $\hat x_j(t)=x_j(t_k^j),\forall t\in [t_k^j,t_{k+1}^j)$ as the state estimation of agent $j$, where $t_k^j$ denotes the $k$-th event-triggering instant of agent $j$.
Then, the main task is to codesign controllers and triggering functions based on $\hat x_j(t)$.

Defining $\hat\xi_i(t)=\sum_{j=1}^Na_{ij}(\hat x_i(t)-\hat x_j(t))$, we first design the following distributed node-based adaptive event-triggered controller for each agent
\begin{equation}\label{controller1}
\begin{aligned}
&u_i(t)=\hat d_i(t)K\hat\xi_i(t),\\
&\dot{d}_i(t)=\hat\xi_i^T(t)\Gamma\hat\xi_i(t),
\end{aligned}
\end{equation}
where $\hat d_i(t)=d_i(t_k^i)$, $\forall t\in [t_k^i,t_{k+1}^i)$, $d_i(t)$ is the adaptive coupling gain for agent $i$ with initial value $d_i(0)\geq 1$, $K$ and $\Gamma$ are feedback gain matrices to be designed.

The key now is to determine when to carry out communications and update controllers.
Define $\tilde x_i(t)=\hat x_i(t)-x_i(t)$ and $\tilde d_i(t)=\hat d_i(t)-d_i(t)$ as the estimation errors.
We design a triggering function for each agent as follows
\begin{equation}\label{trigger}
\begin{aligned}
t_{k+1}^i=\inf\{t>t_k^i~|~\tilde x_i^T(t)&\Gamma \tilde x_i(t)\geq\gamma_i\varepsilon_i(t)\\
&~\vee~|\tilde d_i|\geq \theta_1 e^{-\theta_2(t-t_k^i)}\},
\end{aligned}
\end{equation}
where $t_0^i=0$, $\gamma_i>0$, $\theta_1$, and $\theta_2$ are positive constants, $\varepsilon_i(t)$ is an internal variable with dynamics
\begin{equation}\label{varepsilon}
\begin{aligned}
\dot{\varepsilon}_i(t)=-k_i\varepsilon_i(t)-\sigma_i\tilde x_i^T(t)\Gamma \tilde x_i(t),
\end{aligned}
\end{equation}
with $\varepsilon_i(0)>0$, $k_i>0$, and $\sigma_i>0$.
Only at event instant $t_k^i$, agent $i$ sends its current state information to its out-neighbors. According to \dref{controller1}, agent $i$ updates its control input $u_i$  only when $i$ or its neighboring agents are triggered.

\begin{remark}
It is not difficult to see from \dref{controller1} that the control input $u_i$ is designed completely based on the information at event instants and only updated at some discrete times.
To determine event instants, we propose the novel dynamic rule composed of \dref{trigger} and \dref{varepsilon}, from which, intuitively, the error $\tilde x_i$ and $\tilde d_i$ will not exceed the designed thresholds.
Particularly, the so-called state-based threshold $-\sigma_i \tilde x_i^T(t)\Gamma \tilde x_i(t)$ only relies on agent $i$ and is decoupled from the topology.
With this decoupling feature,  both continuous communications and control updating are avoided.
\end{remark}

Letting $\tilde\xi_i=\sum_{j=1}^Na_{ij}(\tilde x_i-\tilde x_j)$, we have $\tilde\xi_i=\hat\xi_i-\xi_i$. Denote $\bar d_i$ the virtual adaptive gain, whose dynamics are given by
$$
\dot{\bar{d}}_i=\xi_i^T\Gamma\xi_i,
$$
with initial value $\bar d_i(0)\geq 1$. Further let $\check d_i=d_i-\bar d_i$. Then we can derive the closed-loop dynamics of $x_i$ by
\begin{equation*}\label{dxi1}
\begin{aligned}
\dot{x}_i=Ax_i+\bar d_iBK\xi_i+d_iBK\tilde\xi_i+\check d_iBK\xi_i+\tilde d_iBK\hat \xi_i.
\end{aligned}
\end{equation*}
The compact form of the dynamics of $x$ can be written as
\begin{equation}\label{dx1}
\begin{aligned}
\dot{x}=&(I_N\otimes A)x+(\bar D\otimes BK)\xi+(D\otimes BK)\tilde\xi\\
&+(\check D\otimes BK)\xi+(\tilde D\otimes BK)\hat \xi,
\end{aligned}
\end{equation}
where $\xi=[\xi_1^T,\cdots,\xi_N^T]^T$, $\tilde\xi=[\tilde\xi_1^T,\cdots,\tilde\xi_N^T]^T$, $\hat\xi=[\hat\xi_1^T,\cdots,\hat\xi_N^T]^T$, $\bar D=\text{diag}(\bar d_1,\cdots,\bar d_N)$,
$D=\text{diag}(d_1,\cdots,d_N)$, $\tilde D=\text{diag}(\tilde d_1,\cdots,\tilde d_N)$, $\check D=\text{diag}(\check d_1,\cdots,\check d_N)$.

Now, we give the following result and show the convergence of consensus error $\xi$.
\begin{thm}\label{thm1}
For the multi-agent system \dref{model1} under an undirected and connected graph, consensus can be achieved under the distributed adaptive event-triggered protocol \dref{controller1} and \dref{trigger} by choosing $K=-B^TQ$ and $\Gamma=QBB^TQ$ with $Q>0$ whose inverse $P=Q^{-1}$ is the solution to the following LMI:
\begin{equation}\label{lmi1}
AP+PA^T-BB^T<0.
\end{equation}
\end{thm}

\begin{pf}
Consider the Lyapunov function candidate
\begin{equation}\label{lya1}
\begin{aligned}
V_1=x^T(\mathcal{L}\otimes Q)x+\sum_{i=1}^N\left[\frac{1}{2}(\bar d_i-\alpha)^2
+\beta\bar d_iV_0\right],
\end{aligned}
\end{equation}
where $\alpha$ and $\beta$ are positive constants to be determined, and
\begin{equation}\label{V0}
V_0=\sum_{j=1}^N\frac{\varepsilon_j}{k_j}.
\end{equation}
By \dref{varepsilon}, we have $-(k_i+\sigma_i\gamma_i)\varepsilon_i\leq\dot{\varepsilon}_i\leq-k_i\varepsilon_i$, and thus $e^{-(k_i+\sigma_i\gamma_i)t}\varepsilon_i(0)\leq \varepsilon_i(t)\leq e^{-k_it}\varepsilon_i(0)$. It is not difficult to verify that $V_1$ is positive definite. The time derivative of $V_1$ along the trajectory of \dref{dx1} is given by
\begin{equation}\label{dlya11}
\begin{aligned}
\dot{V}_1&=x^T[\mathcal{L}\otimes (QA+A^TQ)]x-2\xi^T(\bar D\otimes \Gamma)\xi\\
&\quad-2\xi^T(D\otimes\Gamma)\tilde\xi-2\xi^T(\check D\otimes\Gamma)\xi-2\xi^T(\tilde D\otimes \Gamma)\tilde \xi\\
&\quad+\sum_{i=1}^N(\bar d_i-\alpha+\beta V_0)\xi_i^T\Gamma\xi_i\\
&\quad-\sum_{i=1}^N\beta\bar d_i\sum_{j=1}^N\left({\varepsilon_j}+\frac{\sigma_j}{k_j}\tilde x_j^T\Gamma \tilde x_j\right).
\end{aligned}
\end{equation}

It is to noted that the key is to analyze the three terms of \dref{dlya11}, i.e., $-2\xi^T(D\otimes\Gamma)\tilde\xi$, $-2\xi^T(\check D\otimes\Gamma)\xi$, and $-2\xi^T(\tilde D\otimes \Gamma)\tilde \xi$.
According to Lemmas \ref{thm1-lemma1}-\ref{thm1-lemma3} in Appendix A, we can obtain that
\begin{equation}\label{dlya12}
\begin{aligned}
\dot V_1
&\leq
x^T[\mathcal L\otimes (AQ+A^TQ)]x+\xi^T[(\beta_2+3\theta_1)I_N\otimes \Gamma]\xi\\
&\quad+\sum_{i=1}^N(-\alpha+\beta V_0)\xi_i^T\Gamma \xi_i-\sum_{j=1}^N\beta \bar d_i\sum_{j=1}^N(\varepsilon_j+\frac{\sigma_j}{k_j}\tilde x_j^T\Gamma \tilde x_j)\\
&\quad+\sum_{i=1}^N(8\bar d_i+2\beta_1+4\theta_1)l_{ii}^2\sum_{j=1}^N\tilde x_j^T\Gamma \tilde x_j,
\end{aligned}
\end{equation}
where $\beta_1$ and $\beta_2$ are defined in Lemmas \ref{thm1-lemma1}-\ref{thm1-lemma3}.

By choosing $\beta\geq2(4+\beta_1+2\theta_1)\max\{l_{ii}^2\}\max\{\frac{k_i}{\sigma_i}\}$,
$\alpha=\alpha_1+\beta V_0(0)+\beta_2+3\theta_1$,  we can obtain from \dref{dlya12} that
\begin{equation}\label{dlya13}
\begin{aligned}
\dot{V}_1&\leq x^T[\mathcal{L}\otimes (QA+A^TQ)]x-\xi^T(\alpha_1I_N\otimes\Gamma)\xi\\
&\leq x^T[\mathcal{L}\otimes (QA+A^TQ-\Gamma)]x\\
&\leq\frac{1}{\lambda_N(\mathcal{L})}\xi^T[I_N\otimes(QA+A^TQ-\Gamma)]\xi\\
&\leq0,
\end{aligned}
\end{equation}
where the second inequality is obtained by choosing $\alpha_1\geq\frac{1}{\lambda_2(\mathcal{L})}$, the last inequality is derived by LMI \dref{lmi1}, $\lambda_2(\mathcal{L})$ and $\lambda_N(\mathcal{L})$ denote the smallest and largest nonzero eigenvalue of $\mathcal{L}$, respectively.

Therefore, $V_1(t)$ is bounded, and so are $\xi_i$ and $\bar d_i$. And $V_1$ has a finite limit as $t\rightarrow\infty$, denoted by $V_1^\infty$.
Integrating the third inequality of \dref{dlya13} yields
$$
\int_0^\infty\xi^T[I_N\otimes(-(QA+A^TQ-\Gamma))]\xi dt\leq\lambda_N(\mathcal{L})(V_1(0)-V_1^\infty).
$$
On the other hand, by \dref{tildedi0} we can get that $d_i=\check{d}_i+\bar d_i\leq2\bar d_i-\bar d_i(0)+d_i(0)+4l_{ii}^2\sum_{j=1}^N \frac{\gamma_j\varepsilon_j(0)}{k_j}$, implying that $d_i$ is bounded.
Then, $\check d_i=d_i-\bar d_i$ is also bounded.
According to the triggering rule, $\tilde x_i^T\Gamma \tilde x_i$ is bounded, implying that $(I_N\otimes BK)\tilde x$ is bounded.
Following this, we have $(I_N\otimes BK)\tilde \xi$ is bounded, further implying that $(I_N\otimes BK)\hat \xi=(I_N\otimes BK)(\xi+\tilde \xi)$ is also bounded.
According to \dref{dx1}, we have
\begin{equation}\label{dxi1'}
\begin{aligned}
\dot{\xi}=&(I_N\otimes A)\xi+(\mathcal L\bar D\otimes BK)\xi+(\mathcal LD\otimes BK)\tilde\xi\\
&+(\mathcal L\check D\otimes BK)\xi+(\mathcal L\tilde D\otimes BK)\hat \xi,
\end{aligned}
\end{equation}
which is bounded.
By Barbalat's lemma \cite{PIoannou1996robust}, we can conclude that $\xi^T[I_N\otimes(-(QA+A^TQ-\Gamma))]\xi\rightarrow0$ as $t\rightarrow\infty$. In this sense, $\xi\rightarrow0$ as $t\rightarrow\infty$, i.e., consensus is achieved.
$\hfill $$\blacksquare$
\end{pf}

\begin{remark}
Note that to avoid continuous control updating, we use discrete value $\hat d_i(t)$ to design the control input $u_i(t)$. Due to its discontinuity, $\hat d_i(t)$ cannot be directly used to establish the Lyapunov function.
To circumvent this difficulty, the virtual adaptive gains $\bar d_i$ are introduced to design the Lyapunov function $V_1$ given in \dref{lya1}.
Furthermore, we also introduce the function $V_0$ into the Lyapunov function $V_1$, in certain sense similar to the idea of
\cite{YuezuLv2020Automatica}, and the boundedness of $V_0$ faciliates the proof of Theorem \ref{thm1}.
\end{remark}


The following result shows that Zeno behavior does not happen.
\begin{thm}\label{thm2}
Under the conditions in Theorem \ref{thm1}, the Zeno behavior is excluded.
\end{thm}

\begin{pf}
We rule out the Zeno behavior by providing a contradiction argument.
Without loss of generality, suppose agent $i$ exists Zeno behavior, i.e., there exists $T<+\infty$ such that $\lim_{k\rightarrow \infty}t_k^i=T$.
Then, for any positive constant $\Delta$, there exists a positive integer $k_0$ such that for $k\geq k_0$, $t_k^i\in[T-\Delta,T)$.

According to Theorem \ref{thm1}, it is not difficult to find that $x_i$, $\hat\xi_i$ and $\hat d_i$ are bounded for time during $[0,T)$.
Noting that $\tilde x_i(t_{k_0}^i)=0$ and $\tilde d_i(t_{k_0}^i)=0$, we have
\begin{equation}\label{com2}
\begin{aligned}
&\tilde x_i^T(t)\Gamma\tilde x_i(t)=\left[\int_{t_{k_0}^i}^t Ax_i+BK\hat d_i\hat \xi_i(\tau)d\tau\right]^T\Gamma \cdot\\
&~~~\left[\int_{t_{k_0}^i}^t Ax_i+BK\hat d_i\hat\xi_i(\tau)d\tau\right]\leq \|\Gamma\|h_1^2(t-t_{k_0}^i)^2,\\
&|\tilde d_i|=\left|\int_{t_{k_0}^i}^t\hat \xi_i^T(\tau)\Gamma\hat\xi_i(\tau)d\tau\right|\leq h_2(t-t_{k_0}^i),
\end{aligned}
\end{equation}
where $h_1$ and $h_2$ denote the upper bounds of $\|Ax_i+BK\hat d_i\hat \xi_i(t)\|$ and $\hat \xi_i^T(t)\Gamma\hat\xi_i(t)$ for $t\in[0,T]$, respectively.

Take $\Delta\triangleq \min\{\Delta_1,\Delta_2\}$, where $\Delta_1=\sqrt{\frac{\gamma_ie^{-(k_i+\sigma_i\gamma_i)T}}{\|\Gamma\|h_1^2}}$ and $\Delta_2=\frac{\theta_1e^{-\theta_2T}}{h_2}$.
According to the triggering rule \dref{trigger}, the next event for agent $i$ after $t_{k_0}^i$ happens if
\begin{equation}\label{com1}
\begin{aligned}
&\tilde x_i(t)\Gamma \tilde x_i(t)\geq\gamma_i\varepsilon_i(t)\geq \gamma_ie^{-(k_i+\sigma_i\gamma_i)T}\\
&~{\rm or}~|\tilde d_i(t)|\geq \theta_1e^{-\theta_2(t-t_k^i)}\geq \theta_1e^{-\theta_2T}.
\end{aligned}
\end{equation}
Then, we can obtain that $t_{k_0+1}^i-t_{k_0}^i\geq \Delta$.
Thus, we have $t_{k_0+1}^i\geq T$, which does not consist with $t_{{k_0}+1}^i<T$.
Therefore, we conclude that there does not exist the Zeno behavior.
$\hfill $$\blacksquare$
\end{pf}

\begin{remark}
Theorems \ref{thm1} and \ref{thm2} show that the presented event-triggered protocol ensures the achievement of consensus and the exclusion of the Zeno behavior.
It is not difficult to see that this protocol does not require any global information of the network and thus is fully distributed.
By using the event triggeringtechnique, we successfully avoid not only continuous communications among neighboring agents but also continuous updating of control inputs.
These features allow the designed protocol to well solve the FDCDCC problem in this paper.
\end{remark}

\begin{remark}
Different from the adaptive event-triggered protocols in \cite{NaHuang2020IS,BCheng2018fully,YangyangQian2019TAC,XianweiLi2020,XianweiLi2020Automatica}, in the current paper we use the law \dref{trigger} with dynamic triggering functions \dref{varepsilon} to determine event instants. The advantage of the proposed fully distributed event-triggered protocol is that it only requires both discrete communications and control updating.
As a comparison, continuous updating of controllers cannot be avoided in \cite{BCheng2018fully,YangyangQian2019TAC,XianweiLi2020,XianweiLi2020Automatica}, continuous communications cannot be avoided in \cite{NaHuang2020IS}, and the event-trigged protocols in
\cite{PYu2018TAC} require global information.
Also notice that in \cite{BCheng2018fully} time-varying gains are designed for edges, which depends on the symmetric feature of the topology.
In other words, it is hard to study general directed graphs under the framework of \cite{BCheng2018fully}.
Instead, in this paper we design time-varying gains for nodes.
Such a node-based design idea provides an opportunity to solve the FDCDCC problem under directed graphs we are going to study in the following section.
Furthermore, with a consistent node-based feature in form, i.e., both the time-varying gains and the triggering functions are designed for nodes, it is clear to use $\hat d_i(t)$ rather than $d_i(t)$ in \dref{controller1} in order to avoid continuous updating.
\end{remark}


\section{FDCDCC for Directed Graphs}\label{s4}
In this section, we will further extend the node-based adaptive event-triggered protocol \dref{controller1} into the protocols applicable to general directed graphs.
It should be pointed out that the Laplacian matrix associated with directed graph is asymmetric, which makes the Lyapunov function \dref{lya1} no longer be positive-definite. For this reason, the protocols designed for undirected graphs in Section \ref{s3} are not applicable and novel protocols need to be explored in order to solve the FDCDCC problem established in Section \ref{s2}.

\subsection{The Case of Strongly Connected Graphs}
In this subsection, we consider the case of strongly connected graphs.
The key here is to tackle the difficulty caused by the unpleasant asymmetric Laplacian matrix.

An extra quadratic function is introduced for each agent and the controllers are designed as follows:
\begin{equation}\label{controller2}
\begin{aligned}
&u_i(t)=(\hat d_i(t)+\rho_i(t))\mu K\hat\xi_i(t),\\
&\dot{ d}_i(t)=\hat\xi_i(t)^T\Gamma\hat\xi_i(t),
\end{aligned}
\end{equation}
where $\mu\in(0,2)$ is a constant, $\rho_i(t)=\hat\xi_i^T(t)Q\hat\xi_i(t)$ is the quadratic form of $\hat\xi_i(t)$, $d_i(t)$ is the adaptive gain with initial value $ d_i(0)\geq1$, $\hat d_i(t)=d_i(t_k^i)$, and the rest variables are the same as in \dref{controller1}.

To determine triggering instants, we design the triggering condition for each agent as follows
\begin{equation}\label{trigger2}
\begin{aligned}
t_{k+1}^i=\inf\{t>t_k^i~|~&\tilde x_i^T(t) \tilde x_i(t)\geq\gamma_i\varepsilon_i(t)\\
&\vee~|\tilde d_i|\geq \theta_1 e^{-\theta_2(t-t_k^i)}\},
\end{aligned}
\end{equation}
where $t_0^i=0$, $\gamma_i>0$ is a positive constant, $\varepsilon_i(t)$ is an internal variable with dynamics
\begin{equation}\label{varepsilon2}
\begin{aligned}
\dot{\varepsilon}_i(t)=-k_i\varepsilon_i(t)-\sigma_i\tilde x_i^T(t) \tilde x_i(t),
\end{aligned}
\end{equation}
with $\varepsilon_i(0)>0$, $k_i>0$ and $\sigma_i>0$.

Denote $\bar d_i$ the virtual adaptive gain with initial value $\bar d_i(0)\geq1$, whose dynamics are given by
$$\dot{\bar{ d}}_i=\xi_i^T\Gamma\xi_i.
$$
Let $\check d_i= d_i-\bar d_i$, $\bar\rho_i=\xi_i^TQ\xi_i$, and $\check\rho_i=\rho_i-\bar\rho_i$. The closed-loop dynamics of $x_i$ can be given by
\begin{equation*}\label{dxi2}
\begin{aligned}
\dot{x}_i&=Ax_i+\mu(\bar d_i+\bar\rho_i)BK\xi_i+\mu( d_i+\rho_i)BK\tilde\xi_i\\
&\quad+\mu(\check d_i+\check\rho_i)BK\xi_i+\mu\tilde d_iBK\hat \xi_i.
\end{aligned}
\end{equation*}
Then, the dynamics of $\xi$ in compact form can be written as
\begin{equation}\label{xi2}
\begin{aligned}
\dot{\xi}=&[I_N\otimes A+\mu\mathcal{L}(\bar D+\bar\rho)\otimes BK]\xi+[\mu\mathcal{L}(D+\rho)\otimes BK]\tilde\xi\\
&+[\mu\mathcal{L}(\check D+\check\rho)\otimes BK]\xi+[\mu\mathcal L\tilde D\otimes BK]\hat \xi,
\end{aligned}
\end{equation}
where $\rho=\text{diag}(\rho_1,\cdots,\rho_N)$,
$\bar\rho=\text{diag}(\bar\rho_1,\cdots,\bar\rho_N)$,
$\check\rho=\text{diag}(\check\rho_1,\cdots,\check\rho_N)$.

The next result shows the achievement of consensus and the exclusion of the Zeno behavior under strongly connected graphs.
\begin{thm}\label{thm5}
Suppose that the network topology associated with the $N$ agents in \dref{model1} is strongly connected. The consensus can be achieved under the adaptive event-triggered protocol \dref{controller2} and \dref{trigger2}. Moreover, the Zeno behavior is excluded.
\end{thm}

Before giving the proof of Theorem \ref{thm5}, we introduce the following lemma.
\begin{lem}[\cite{JieMei2016TAC}]\label{lem1}
For a strongly connected graph $\mathcal{G}$ with Laplacian matrix $\mathcal L$, the matrix $\hat{\mathcal L}= R\mathcal L+\mathcal L^TR$ represents a weighted symmetric Laplacian matrix of an undirected connected graph, where $R=\text{diag}(r_1,\cdots,r_N)>0$ with $r=[r_1,\cdots,r_N]$ being the left zero unit eigenvector of $\mathcal L$. Moreover, $\min_{z^Tx=0}x^T\hat{\mathcal L}x\geq\frac{\lambda_2(\hat{\mathcal L})}{N}x^Tx$, where $\lambda_2(\hat{\mathcal L})$ is the smallest nonzero eigenvalue of $\hat{\mathcal L}$ and $z$ is a vector with all positive elements.
\end{lem}

Now, we are ready to prove Theorem \ref{thm5}.

\begin{pf}
Consider the Lyapunov function candidate
\begin{equation}\label{lya2}
\begin{aligned}
V_2=&\sum_{i=1}^Nr_i\left((\bar d_i+\frac{1}{2}\bar\rho_i)\bar\rho_i+ \frac{(\bar d_i-\alpha)^2}{2}+\beta\bar d_i^2V_{0}\right),
\end{aligned}
\end{equation}
where $\alpha$ and $\beta$ are positive constants, $r_i$ is defined by Lemma \ref{lem1}, and $V_0$ is defined in \dref{V0}.
The time derivative of $V_2$ is
\begin{equation}\label{dlya21}
\begin{aligned}
\dot{V}_2
&=\xi^T[(\bar D+\bar\rho)R\otimes(QA+A^TQ)\\
&\quad-\mu(\bar D+\bar\rho)(R\mathcal{L}+\mathcal{L}^TR)(\bar D+\bar\rho)\otimes\Gamma]\xi\\
&\quad-2\mu\xi^T[(\bar D+\bar\rho)R\mathcal L(D+\rho)\otimes\Gamma]\tilde\xi\\
&\quad-2\mu\xi^T[(\bar D+\bar\rho)R\mathcal L(\check D+\check\rho)\otimes\Gamma]\xi\\
&\quad-2\mu\xi^T[(\bar D+\bar \rho)R\mathcal L\tilde D\otimes \Gamma]\hat\xi\\
&\quad+\xi^T[((1+2\beta V_0)\bar D+\bar\rho-\alpha I_N)R\otimes\Gamma]\xi\\
&\quad-\sum_{i=1}^N\beta r_i\bar d_i^2\sum_{j=1}^N\left({\varepsilon_j}+\frac{\sigma_j}{k_j}\tilde x_j^T\tilde x_j\right).
\end{aligned}
\end{equation}

In light of Lemma \ref{lem1}, and noticing that
\begin{equation*}
\begin{aligned}
&((\bar D+\bar\rho)^{-1}r^T\otimes\textbf{1}_p)^T((\bar D+\bar\rho)\otimes B^TP)\xi\\
=&(r\mathcal{L}\otimes\textbf{1}_p^T B^TP)x=0,
\end{aligned}
\end{equation*}
we have
\begin{equation}
\begin{aligned}
&-\mu\xi^T[(\bar D+\bar\rho)(R\mathcal L+\mathcal L^TR)(\bar D+\bar\rho)\otimes\Gamma]\xi\\
\leq&-\xi^T\left[\frac{\mu\lambda_2(\hat L)}{N}(\bar D+\bar\rho)^2\otimes\Gamma\right]\xi.
\end{aligned}
\end{equation}

According to Lemmas \ref{thm5-lemma1}-\ref{thm5-lemma3} in Appendix B, for $t>T_0$, where $T_0$ denotes a positive constant,
by choosing $\alpha=\frac{\mu\beta_1^2\sigma_{\max}(R{\mathcal L})}{\alpha_1}+\frac{16N\mu\theta^2}{\lambda_2(\hat {\mathcal L})}\sigma_{\max}(R\mathcal L)
+\alpha_0$ and $\beta\geq\mu\left(\frac{2(1+\alpha_1+\beta_2)^2}{\alpha_1}+\frac{32N\theta^2}{\lambda_2(\hat {\mathcal L})}\right)\frac{\sigma_{\max}(R\mathcal L)
\lambda_{\max}(\Gamma)}{\lambda_{\min}(R)}
\max\{l_{ii}^2\}\max\{\frac{k_i}{\sigma_i}\}$, we can get that
\begin{equation}
\begin{aligned}
\dot{V}_2&\leq\xi^T[(\bar D+\bar\rho)R\otimes(QA+A^TQ+\frac{\mu}{2}X+\beta_0\Gamma)]\xi\\
&\quad-\xi^T[(\frac{\mu\lambda_2(\hat{\mathcal{L}})}{8N}(\bar D+\bar\rho)^2 +\alpha_0R)\otimes\Gamma]\xi,
\end{aligned}
\end{equation}
where $\beta_0=1+2\beta V_0(0)+\frac{2\mu\beta_1\sigma_{\max}(R\mathcal L)}{\lambda_{\min}(R)}$. By choosing $\alpha_0\geq\frac{2N(1+\beta_0)^2\lambda_{\max}(R)}{\mu\lambda_2(\hat{\mathcal{L}})}$ and using the Young's Inequality \cite{DBernstein2009Matrix}, we can derive that for $t>T_0$,
\begin{equation}
\begin{aligned}
\dot{V}_2&\leq\xi^T[(\bar D+\bar\rho)R\otimes(QA+A^TQ+\frac{\mu}{2}X+\beta_0\Gamma)]\xi\\
&\quad-(1+\beta_0)\xi^T[(\bar D+\bar\rho)R\otimes\Gamma]\xi\\
&=-\frac{2-\mu}{2}\xi^T[(\bar D+\bar\rho)R\otimes X]\xi\\
&\leq0.
\end{aligned}
\end{equation}
Therefore, $V_2$ is bounded, and so are $\bar d_i$ and $\bar\rho_i$. Following the similar steps and in light of Barbalat's Lemma, we can conclude that $\xi_i$ asymptotically converges to zero, i.e., the consensus is indeed realized. Taking similar steps in the proof of Theorem \ref{thm2}, we can derive that the Zeno behavior never happens.
$\hfill $$\blacksquare$
\end{pf}

\begin{remark}
It is to be noted that we embed a small positive constant $\mu$ into the controller \dref{controller2} and it plays a role of reducing the amplitude of the control input.  Generally, a small $\mu$ will reduce the control amplitude and meanwhile lower the speed of convergence. So, we need to choose a proper $\mu$ to trade-off the control amplitude and the convergence speed.
\end{remark}

\begin{remark}
It is to be noted that in order to deal with the unpleasant asymmetric Laplacian matrix of directed graphs, we have included an extra quadratic function $\rho_i(t)$ into the designed controller of each agent. This idea is partly motivated by \cite{YuezuLv2016Automatica}, which nevertheless considers continuous control and communications.
\end{remark}

\subsection{The Case of Leader-Follower Directed Graphs}
Note that we have only considered the case of strongly connected graphs in the last subsection.
In this subsection, we aim at considering leader-follower directed graphs which only contains a directed spanning tree.
Without loss of generality, we assume that agent 1 is the leader and the other agents are the followers.

\begin{assum}\label{assum-dst}
The graph associated with the $N$ agents has a directed spanning tree with the leader as a root.
\end{assum}

Under this assumption, the Laplacian matrix $\mathcal L$ can be partitioned as $\mathcal L=\left[\begin{matrix}0&0_{1\times (N-1)}\\\mathcal L_2&\mathcal L_1\end{matrix}\right]$, where $\mathcal L_1$ is a non-singular $M$-matrix.


To track the leader's state, we propose a distributed adaptive event-based controller for each follower as
\begin{equation}\label{pro-ada-direc-spanning}
\begin{aligned}
&u_i(t)=(\hat d_i(t)+\rho_i(t))\mu K\hat\xi_i(t),\\
&\dot{ d}_i(t)=\hat\xi_i^T(t)\Gamma\hat\xi_i(t),~i=2,\cdots,N,
\end{aligned}
\end{equation}
where $\mu\in(0,2)$ is a constant, $\rho_i=\hat\xi_i^TQ\hat\xi_i$ is the quadratic form of $\hat\xi_i$, $d_i$ is the adaptive gain with initial value $ d_i(0)\geq1$, $\hat d_i(t)=d_i(t_k^i)$, and the rest variables are the same as in \dref{controller1}.

Let $\tilde x_i=\hat x_i-x_i$ and $\tilde d_i=\hat d_i-d_i$.
The triggering condition for each agent is designed by
\begin{equation}\label{trigger2-t-st}
\begin{aligned}
t_{k+1}^1&=\inf\{t>t_k^1~|~\tilde x_1^T(t) \tilde x_1(t)\geq\gamma_1\varepsilon_1(t)\},\\
t_{k+1}^i&=\inf\{t>t_k^i~|~\tilde x_i^T(t) \tilde x_i(t)\geq\gamma_i\varepsilon_i(t)\\
&~~~~~\vee~|\tilde d_i|\geq \theta_1 e^{-\theta_2(t-t_k^i)}\},~i=2,\cdots,N,
\end{aligned}
\end{equation}
where $t_k^0=0$, $\gamma_i>0$ is a positive constant, $\varepsilon_i(t)$ is an internal variable with dynamics
\begin{equation}\label{varepsilon3-st}
\begin{aligned}
\dot{\varepsilon}_i(t)=-k_i\varepsilon_i(t)-\sigma_i\tilde x_i^T(t) \tilde x_i(t),~i=1,\cdots,N,
\end{aligned}
\end{equation}
with $\varepsilon_i(0)>0$, $k_i>0$ and $\sigma_i>0$.
At event instants of the leader $t_k^1$, the leader sends its current information to the informed followers.
Differently, when event of followers happens, the corresponding followers need to update their controllers and carry out communications among neighboring followers.

\begin{thm}\label{theorem-adative-spanning}
Suppose that Assumption \ref{assum-dst} holds. Under the adaptive event-triggered protocol composed of \dref{pro-ada-direc-spanning}-\dref{varepsilon3-st}, the leader-follower consensus of systems \dref{model1} can be achieved and the time-varying weights $d_i(t)$ are bounded.
Moreover, there does not exist Zeno behavior for the whole multi-agent systems.
\end{thm}

\begin{remark}
We omit the proof of Theorem \ref{theorem-adative-spanning} here for conciseness and refer interested readers to finish it following the proof of Theorem \ref{thm5}. For the leader-follower network, we design a triggering function for the leader to determine when to transmit information from the leader to informed followers, and triggering functions for followers used to determine when to update controllers and carry out communications among neighboring followers.
\end{remark}

\section{Output-Feedback Based FDCDCC}\label{s5}
In this section, we are going to extend the results of Sections \ref{s3} and \ref{s4} to the case where state information is not available.
In this case, we need to design novel output-feedback control algorithms to solve event-triggered consensus problem.

\subsection{The Case of Undirected Graphs}

We first consider undirected graphs in this subsection.
To deal with the difficulty caused by unavailable state information, the idea here is to design state observer using local output information.
Then,  we use the state observer to design controllers and triggering functions.

Now,  we propose the following adaptive output-based event-triggered controller:
\begin{equation}\label{controller-u-o}
\begin{aligned}
\dot v_i(t)&=Av_i(t)+Bu_i(t)+F(Cv_i(t)-y_i(t)),\\
\dot e_i(t)&=\hat \eta_i^T(t)\Gamma \hat\eta_i(t),\\
u_i(t)&=\hat e_i(t)K\hat\eta_i(t),~i=1,\cdots,N,
\end{aligned}
\end{equation}
where $v_i(t)$ are state observers, $e_i(0)\geq 1$, $\hat e_i(t)=e_i(t_k^i)$, $\hat \eta_i(t)=\sum_{j=1}^Na_{ij}(\hat v_i(t)-\hat v_j(t))$, $\hat v_i=v_i(t_k^i)$, $\forall t\in[t_k^i,t_{k+1}^i)$, $i=1,\cdots,N$.

By defining $\tilde v_i=\hat v_i-v_i$ and $\tilde e_i=\hat e_i-e_i$ as the estimation errors, we design the following triggering condition
\begin{equation}\label{triggering-u-o}
\begin{aligned}
t_{k+1}^i\triangleq \inf\{t>t_k^i~&|~\tilde v_i^T(t)\Gamma \tilde v_i(t)\geq \gamma_i\epsilon_i(t)\\
&~\vee~|\tilde e_i|\geq \theta_1 e^{-\theta_2(t-t_k^i)}\},
\end{aligned}
\end{equation}
where $\epsilon_i(t)$ is an internal variable with dynamics
\begin{equation}\label{epsilon}
\begin{aligned}
\dot{\epsilon}_i(t)=-k_i\epsilon_i(t)-\sigma_i\tilde v_i^T(t)\Gamma \tilde v_i(t),
\end{aligned}
\end{equation}
with $\epsilon_i(0)>0$, and other parameters being the same as in \dref{trigger}.

Let $z_i=v_i-x_i$, $i=1,\cdots,N$. Then, we have
\begin{equation*}
\begin{aligned}
\dot z_i&=(A+FC)z_i,\\
\dot v_i&=Av_i+Bu_i+FCz_i,~i=1,\cdots,N.
\end{aligned}
\end{equation*}

Denote $\eta=[\eta_1^T,\cdots,\eta_N^T]^T=(\mathcal L\otimes I_n)v$.
Letting $\tilde\eta_i=\sum_{j=1}^Na_{ij}(\tilde v_i-\tilde v_j)$, we have $\tilde\eta_i=\hat\eta_i-\eta_i$. Denote $\bar e_i$ the virtual adaptive gain, whose dynamics are given by
$$
\dot{\bar{e}}_i=\eta_i^T\Gamma\eta_i,
$$
with initial value $\bar e_i(0)\geq 1$. Further let $\check e_i=e_i-\bar e_i$. Then we can derive the closed-loop dynamics of $v_i$ by
\begin{equation*}\label{dxi1-o}
\begin{aligned}
\dot{v}_i=Av_i+\bar e_iBK\eta_i+e_iBK\tilde\eta_i+\check e_iBK\eta_i+\tilde e_iBK\hat \eta_i+FCz_i.
\end{aligned}
\end{equation*}
And the compact form of the dynamics of $v$ can be written as
\begin{equation}\label{dx1-o}
\begin{aligned}
\dot{v}=&(I_N\otimes A)v+(\bar E\otimes BK)\eta+(E\otimes BK)\tilde\eta\\
&+(\check E\otimes BK)\eta+(\tilde E\otimes BK)\hat \eta+(I_N\otimes FC)z,
\end{aligned}
\end{equation}
where $\tilde\eta=[\tilde\eta_1^T,\cdots,\tilde\eta_N^T]^T$, $\hat\eta=[\hat\eta_1^T,\cdots,\hat\eta_N^T]^T$, $\bar E=\text{diag}(\bar e_1,\cdots,\bar e_N)$,
$E=\text{diag}(e_1,\cdots,e_N)$, $\tilde E=\text{diag}(\tilde e_1,\cdots,\tilde e_N)$, $\check E=\text{diag}(\check e_1,\cdots,\check e_N)$, and $z=[z_1^T,\cdots,z_N^T]^T$.

Now,  we introduce the result under the proposed adaptive event-triggered output-feedback protocol.
\begin{thm}\label{thm-o}
Choose $F$ such that $A+FC$ is Hurwitz and choose $K=-B^TQ$ and $\Gamma=QBB^TQ$ where $Q>0$ is the solution to the following Riccati equation:
\begin{equation}\label{lmi1-o}
A^TQ+QA-QBB^TQ+I=0.
\end{equation}
Then, all agents can achieve consensus and do not exist Zeno behavior.
\end{thm}

\begin{pf}
Consider the Lyapunov function candidate
\begin{equation}\label{lya1-o}
\begin{aligned}
V_3=v^T(\mathcal{L}\otimes Q)v+\sum_{i=1}^N\left[\frac{1}{2}(\bar e_i-\alpha)^2
+\beta\bar d_iV_{\epsilon}+\gamma z_i^TQ_0z_i\right],
\end{aligned}
\end{equation}
where $\alpha$, $\beta$, and $\gamma$ are positive constants, and $V_{\epsilon}=\sum_{j=1}^N\frac{\epsilon_j}{k_j}$, and $Q_0$ is a positive definite matrix such that $Q_0(A+FC)+(A+FC)^TQ_0<0$.

When computing the time derivate of $V_3$, we need to note that
\begin{equation*}
\begin{aligned}
2v^T(\mathcal L\otimes Q)&(I_N\otimes FC)z\leq \frac{1}{2}v^T(\mathcal L\otimes I_n)v\\
&+2\lambda_N(\mathcal L)\lambda_{\max}(C^TF^TQ^2FC)z^Tz,
\end{aligned}
\end{equation*}
and
\begin{equation*}
\begin{aligned}
\frac{d z_i^TQ_0z_i}{dt}&=z_i^T[Q_0(A+FC)+(A+FC)^TQ_0]z_i\\
&\leq -\lambda_{\min}(Q_0(A+FC)+(A+FC)^TQ_0)z_i^Tz_i.
\end{aligned}
\end{equation*}

Then, following the steps and choosing $\alpha$ and $\beta$ as in the proof of Theorem \ref{thm1}, we let $\gamma\geq \frac{2\lambda_N(\mathcal L)\lambda_{\max}(C^TF^TQ^2FC)}{\lambda_{\min}(Q_0(A+FC)+(A+FC)^TQ_0)}$ and conclude that
\begin{equation}\label{dlya13-o}
\begin{aligned}
\dot{V}_3&\leq x^T[\mathcal{L}\otimes (QA+A^TQ-\Gamma+\frac{1}{2}I)]x\\
&\leq\frac{1}{\lambda_N(\mathcal{L})}\xi^T[I_N\otimes(QA+A^TQ-\Gamma+\frac{1}{2}I)]\xi\\
&\leq 0.
\end{aligned}
\end{equation}
Then, we can show that $v_i$ asymptotically converge to be the same, further implying that consensus can be achieved.
The Zeno behavior can be excluded as in the proof of Theorem \ref{thm2}. The details are skipped for brevity.
$\hfill $$\blacksquare$
\end{pf}

\begin{remark}
To avoid using the state information,  in this subsection we propose a kind of adaptive observer-based event-triggered protocol \dref{controller-u-o}.
It is to be noted that the design of the observer is not unique and other kinds of observers may also work.
To be convenient, we choose the design matrix $Q$ by solving the Riccati equation \dref{lmi1-o}.
Actually, when the pair $(A,B)$ is stabilizable, the solvability of the Riccati equation \dref{lmi1-o} is equal to that of the LMI \dref{lmi1} \cite{ZLi2014cooperative}.
\end{remark}

\subsection{The Case of Directed Graphs}
In this subsection, we consider strongly connected graphs and extend to the results of Section IV-A to the case where only local output information is available.

We propose the following adaptive output-based event-triggered controller for each agent:
\begin{equation}\label{controller-u-o-d}
\begin{aligned}
\dot v_i(t)&=Av_i(t)+Bu_i(t)+F(Cv_i(t)-y_i(t)),\\
\dot e_i(t)&=\hat \eta_i^T(t)\Gamma \hat\eta_i(t),\\
u_i(t)&=(\hat e_i(t)+\varrho_i(t))\mu K\hat\eta_i(t),~i=1,\cdots,N,
\end{aligned}
\end{equation}
where $\mu\in(0,2)$ is a constant, $e_i(0)\geq 1$, $\hat e_i(t)=e_i(t_k^i)$, $\hat \eta_i(t)=\sum_{j=1}^Na_{ij}(\hat v_i(t)-\hat v_j(t)$), $\hat v_i(t)=v_i(t_k^i)$, $\varrho_i=\hat\eta_i^TQ\hat\eta_i$ is the quadratic form of $\hat\eta_i$, $\forall t\in[t_k^i,t_{k+1}^i)$, $i=1,\cdots,N$.

To determine event instants, we design the following triggering condition for each agent
\begin{equation}\label{triggering-u-o-d}
\begin{aligned}
t_{k+1}^i\triangleq \inf\{t>t_k^i~&|~\tilde v_i^T(t)\tilde v_i(t)\geq \gamma_i\epsilon_i(t)\\
&~\vee~|\tilde e_i|\geq \theta_1 e^{-\theta_2(t-t_k^i)}\},
\end{aligned}
\end{equation}
where $\epsilon_i(t)$ is an internal variable with dynamics
\begin{equation}\label{epsilon-o}
\begin{aligned}
\dot{\epsilon}_i(t)=-k_i\epsilon_i(t)-\sigma_i\tilde v_i^T(t)\tilde v_i(t),
\end{aligned}
\end{equation}
with $\epsilon_i(0)>0$, and other parameters being the same as in \dref{trigger}.

\begin{thm}\label{thm-o-d}
Choosing feedback matrices $F$, $K$, and $\Gamma$ as in Theorem \ref{thm-o}, the consensus of \dref{model1} can be achieved and the time-varying weights $e_i(t)$ are bounded. Moreover, there does not exist Zeno behavior for the whole multi-agent systems.
\end{thm}

\begin{remark}
Theorem \ref{thm-o-d} can be proved following the steps as in the proof of Theorems \ref{thm5} and \ref{thm-o}.
Here we omit the details for brevity.
Under the adaptive output-feedback event-triggered protocol composed of \dref{controller-u-o-d} and \dref{triggering-u-o-d}, the observer $v_i$ will asymptotically converge to the state $x_i$.
\end{remark}

%

\section{Simulation Examples and Discussions}\label{s6}
\begin{figure}[htbp]
\centering
\includegraphics[width=0.22\textheight,height=0.22\textwidth]{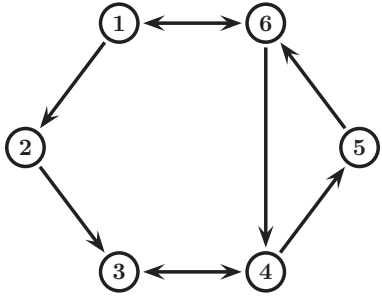}
\caption{The network topology associated with the agents.}
\label{fig-top}
\end{figure}

In this section we use the following numerical example to show the effectiveness of the proposed fully distributed event-triggered protocols.
Particularly, we consider six agents whose dynamics satisfy \dref{model1} with $A=\left[\begin{smallmatrix}0&1&0\\-1&0&0\\0&0&0.1\end{smallmatrix}\right]$ and $B=\left[\begin{smallmatrix}0\\1\\1\end{smallmatrix}\right]$.
The six agents are networked by the graph shown in Fig. \ref{fig-top}, which is strongly connected.

To achieve consensus of the above problem, we use the adaptive event-triggered protocol \dref{controller2} and \dref{trigger2}.
Solving the LMI \dref{lmi1} gives $P=\left[\begin{smallmatrix}0.4792&-0.1955&0.3510\\0.1955&1.0261&0.7196\\0.3510&0.7196&1.0700\end{smallmatrix}\right]$.
Then, we obtain the feedback matrices $K=\left[\begin{smallmatrix}0.8409&-0.5412&-0.8465\end{smallmatrix}\right]$ and $\Gamma=\left[\begin{smallmatrix}0.7071&-0.4551&-0.7118\\-0.4551&0.2929&0.4581\\-0.7118&0.4581&0.7165\end{smallmatrix}\right]$.
The rest parameters in the protocol are selected as
$\mu=1$, $\gamma_i=1$, $\theta_1=1$, $\theta_2=1$, $k_i=0.25$, $\sigma_i=0.25$, and $\varepsilon_i(0)=0.4$, $i=1,\cdots,6$.

\begin{figure}[htbp]
\centering
\includegraphics[scale=1]{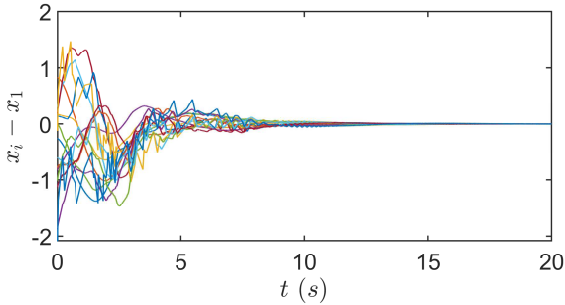}
\caption{Consensus error $x_i-x_1$, $i=2,\cdots,6$.}
\label{fig-states}
\end{figure}

\begin{figure}[htbp]
\centering
\includegraphics[scale=1]{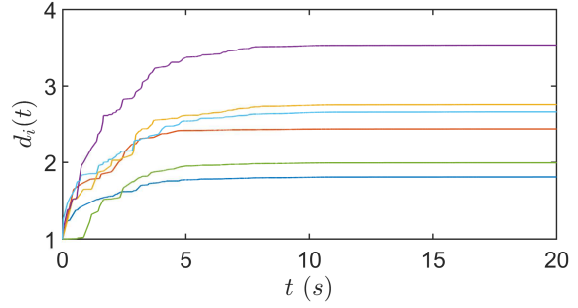}
\caption{The time-varying gains $d_i$, $i=1,\cdots,6$.}
\label{fig-c}
\end{figure}

\begin{figure}[htbp]
\centering
\includegraphics[scale=1]{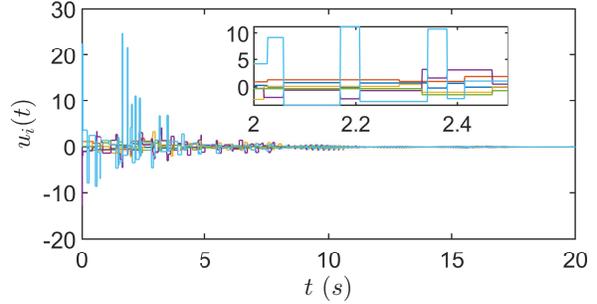}
\caption{Control input $u_i$ of each agent.}
\label{fig-u}
\end{figure}

\begin{figure}[htbp]
\centering
\includegraphics[scale=1]{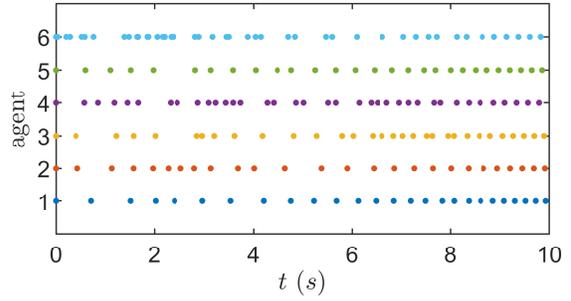}
\caption{Triggering instants of each agent.}
\label{fig-tri-ins}
\end{figure}

Running the software named MATLAB in a computing frequency of 1 GHz, we can get the following results. Fig. \ref{fig-states} shows that consensus can be achieved. The time-varying gains $d_i$ are shown in Fig. \ref{fig-c}, implying that $d_i$ tend to fixed values.
According to Fig. \ref{fig-u}, control input is updated at event instants and is through zero-order hold until next event instant.
In this simulation, communications for each agent are only needed at the discrete instants shown in Fig. \ref{fig-tri-ins}.


In Sections \ref{s3}-\ref{s5}, we have proposed a series of fully distributed adaptive event-triggered protocols with discrete communication and updating.
The core is to design a dynamic triggering condition for each agent and only it is triggered, the agent sends its current state or output information to its out-neighbors.
Naturally, there is a different idea that only when the triggering condition is triggered, the agent samples information from its neighbors and updates its controller. Following this idea, the event-triggered protocols with relative state measurement information are studied in \cite{XianweiLi2020,XianweiLi2020Automatica}.

Noting that the controllers of \cite{XianweiLi2020,XianweiLi2020Automatica} are required to be continuously updated, we in \cite{BCheng2020CDC} design an adaptive event-triggered controller based on relative state measurement as follows
\begin{equation}\label{pro-ada-dis-direc}
\begin{aligned}
u_i(t)&=(c_i(t_k^i)+\rho_i(t_k^i))K\xi_i(t_k^i),\\
\dot c_i(t)&=\kappa_i\xi_i^T(t_k^i)\Gamma\xi_i(t_k^i),
\end{aligned}
\end{equation}
where $\rho_i(t)=\xi_i^T(t)Q\xi_i(t)$, $c_i(0)$ and $\kappa_i$ are positive constants, $Q>0$ and $\Gamma$ are constant matrices, $\xi_i(t)=\sum_{j=1}^Na_{ij}(x_i(t)-x_j(t))$.
Denote $\hat c_i(t)\triangleq c_i(t_k^i)$, $\hat \rho_i(t)\triangleq \rho_i(t_k^i)$.
Triggering functions are designed as follows
\begin{equation}\label{tri-ada-direc}
\begin{aligned}
f_{1i}&=(c_i+\rho_i)^2\tilde\xi_i^T\Gamma\tilde\xi_i-\phi_1\xi_i^T\Gamma\xi_i-\mu_{1i}(t),\\
f_{2i}&=(\tilde c_i+\tilde \rho_i)^2\hat\xi_i^T\Gamma\hat\xi_i-\phi_2\xi_i^T\Gamma\xi_i-\mu_{2i}(t),
\end{aligned}
\end{equation}
where $\tilde c_i=\hat c_i-c_i$, $\tilde\rho_i=\hat\rho_i-\rho_i$, $\hat\xi_i(t)=\xi_i(t_k^i)$, $\phi_1$ and $\phi_2$ are positive constants, and $\mu_{1i}(t)$ and $\mu_{2i}(t)$ are positive scalar functions belonging to $L_1[0,+\infty)$, indicating that $L_1[0,+\infty)\triangleq \{x(t)\in \mathbf R|\int_0^\infty|x(\tau)|d\tau<\infty\}$.
Exactly, $\mu e^{-\nu t}$ can be a special case of $\mu_{1i}(t)$.
We define $t_{k+1}^i\triangleq \inf\{t>t_k^i~|~f_{1i}(t)\geq 0~\vee~f_{2i}(t)\geq 0\}$.
Once $f_{1i}\geq 0$ or $f_{2i}\geq 0$, agent $i$ immediately samples information from its neighbors and updates its control input $u_i(t)$. At other times, $u_i(t)$ keeps zero-order hold.

\begin{lem}[\cite{BCheng2020CDC}]\label{theorem-adative-strongly}
Letting $c_i(0)>\sqrt{2\phi_1}$, under the adaptive event-triggered protocol composed of \dref{pro-ada-dis-direc} and \dref{tri-ada-direc} with $K=-B^TQ$ and $\Gamma=QBB^TQ$, where $Q>0$ is the solution to the Riccati equation \dref{lmi1-o}, consensus of systems \dref{model1} can be achieved and the time-varying weights $c_i(t)$ are bounded.
Moreover, there does not exist Zeno behavior for the whole multi-agent systems.
\end{lem}


Note that the triggering function \dref{tri-ada-direc} relies on continuous relative state information $\xi_i(t)$ of neighboring agents.
Even if as a compensation we can compute $\xi_i(t)$ according to the dynamics $\dot \xi_i(t)=A\xi_i(t)+B\sum_{j=1}^Na_{ij}(u_i(t)-u_j(t))$, each agent needs to use the inputs of neighbors as pointed out in \cite{BCheng2020CDC,Cheng2016Event,YangyangQian2019TAC}.
This means that each agent needs to transmit its input value to its out-neighbors at event instants, which takes the same way of the information interaction with the fully distributed adaptive event-triggered protocols in Sections \ref{s3}-\ref{s5}.
In addition, the computing of the external variable $\xi_i(t)$ also increases computational cost.
Furthermore, the reliability of the calculated consensus error $\xi_i$ depends on how accurate the dynamics are. Note that once there exist unmodeled dynamics or disturbances, $\xi_i(t)$ cannot be accurately calculated by the above computing equation anymore.
These drawbacks essentially exist in the proposed protocol \dref{pro-ada-dis-direc} as well as the existing protocols based on relative state measurements, such as the ones of \cite{XianweiLi2020,XianweiLi2020Automatica}.
Nevertheless, such limitations can be circumvented by using the distributed adaptive dynamic event-triggered protocols proposed in Sections \ref{s3}-\ref{s5}.

\begin{table}
\caption{Triggering numbers of agents under \dref{controller2} and \dref{pro-ada-dis-direc}.}
\centering
\begin{tabular}{ccccccc}  
\hline
Agent &1 &2 &3 &4 &5 &6\\ \hline
under \dref{controller2} &219 &203 &220 &212 &214 &223\\
under \dref{pro-ada-dis-direc} &63 &78 &72 &81 &49 &128\\
\hline
\end{tabular}
\end{table}
We also do the simulation using the protocol \dref{pro-ada-dis-direc} and \dref{tri-ada-direc}.
In the simulation, we choose parameters as: $\kappa_i=2$, $\phi_1=\phi_2=1$, $\mu_{1i}(t)=\mu_{2i}(t)=e^{-0.25t}$, $c_i(0)=2$.
We count the number of events for time during $[0,20]s$ for the above two methods as in Table I.
According to Table I, we find that the number of events under the adaptive dynamic event-triggered protocol \dref{controller2} and \dref{trigger2} is more than the one under \dref{pro-ada-dis-direc} and \dref{tri-ada-direc}. This result is not surprising because the protocol with relative state measurement needs to pay the cost of using continuous relative state information among neighboring agents.

\section{Conclusion}\label{s7}
This paper has considered the distributed consensus control problem with the constraints of communication bandwidth and updating frequency.
We have established a novel framework and systematically studied both undirected graph and directed graph, and designed several kinds of adaptive event-triggered protocols.
It has been shown that the presented protocols can guarantee the achievement of the expected control performance and reduce the frequencies of communication among neighboring agents and control updating.
Furthermore, the proposed protocols are fully distributed and scalable, without requiring any global information of the whole network.
The effectiveness of the proposed protocols are clearly verified by the simulation examples and the comparisons with others protocols given in existing related papers.
The results of the current paper can be further extended to several problems, such as tracking control with a dynamic leader whose input is unknown, output regulation of heterogeneous multi-agent systems, and formation control.

\section*{Appendix A. The lemmas in Section \ref{s3}}
For the network dynamics \dref{dx1}, we can derive the following lemmas.
\begin{lem}\label{thm1-lemma1}
It holds that
\begin{equation}
\begin{aligned}
-2\xi^T(D\otimes\Gamma)\tilde\xi
&\leq\frac{1}{2}\xi^T(\bar D\otimes\Gamma)\xi+\frac{1}{2}\xi^T(\check D\otimes\Gamma)\xi\\
&\quad+\sum_{i=1}^N\left(8\bar d_i+2\beta_1\right)l_{ii}^2\sum_{j=1}^N\tilde x_j^T\Gamma \tilde x_j,
\end{aligned}
\end{equation}
 where $\beta_1\geq\max\{-2\bar d_i(0)+2\check d_i(0)+8l_{ii}^2\sum_{j=1}^N \frac{\gamma_j\varepsilon_j(0)}{k_j}\}$ is a positive constant.
\end{lem}

\begin{pf}
By the Young's inequality \cite{DBernstein2009Matrix}, we have
\begin{equation*}\label{xieta}
\begin{aligned}
&\quad-2\xi^T(D\otimes\Gamma)\tilde\xi
\leq\frac{1}{2}\xi^T(D\otimes\Gamma)\xi+2\tilde\xi^T(D\otimes\Gamma)\tilde\xi\\
&=\frac{1}{2}\xi^T(\bar D\otimes\Gamma)\xi+\frac{1}{2}\xi^T(\check D\otimes\Gamma)\xi+2\tilde\xi^T(\bar D\otimes\Gamma)\tilde\xi\\
&\quad+2\tilde\xi^T(\check D\otimes\Gamma)\tilde\xi.
\end{aligned}
\end{equation*}
Note that
\begin{equation}\label{tilded0}
\begin{aligned}
&\check{d}_i(t)=\int_{0}^t(\hat\xi_i^T(\tau)\Gamma\hat\xi_i(\tau)-\xi_i^T(\tau)\Gamma\xi_i(\tau))d\tau+\check d_i(0)\\
&=\int_{0}^t(\tilde\xi_i^T(\tau)\Gamma\tilde\xi_i(\tau)+2\xi_i^T(\tau)\Gamma\tilde\xi_i(\tau))d\tau+\check d_i(0)\\
&\leq\int_0^t\xi_i^T(\tau)\Gamma\xi_i(\tau)d\tau
+2\int_0^t\tilde\xi_i^T(\tau)\Gamma\tilde\xi_i(\tau)d\tau+\check d_i(0)\\
&=\bar d_i(t)+2\int_0^t\tilde\xi_i^T(\tau)\Gamma\tilde\xi_i(\tau)d\tau+\check d_i(0)-\bar d_i(0),
\end{aligned}
\end{equation}
and
\begin{equation}\label{eta}
\begin{aligned}
\tilde\xi_i^T\Gamma\tilde\xi_i
&=\sum_{j=1}^Na_{ij}(\tilde x_i-\tilde x_j)^T\Gamma \sum_{j=1}^Na_{ij}(\tilde x_i-\tilde x_j)\\
&\leq 2l_{ii}^2\tilde x_i^T\Gamma \tilde x_i+2\sum_{j=1}^Na_{ij}\tilde x_j^T\Gamma \sum_{j=1}^Na_{ij}\tilde x_j\\
&\leq 2l_{ii}^2\tilde x_i^T\Gamma \tilde x_i+2l_{ii}\sum_{j=1}^Na_{ij}\tilde x_j^T\Gamma \tilde x_j\\
&\leq 2l_{ii}^2\sum_{j=1}^N\tilde x_j^T\Gamma \tilde x_j.
\end{aligned}
\end{equation}
Combining \dref{tilded0} and \dref{eta}, we can get that
\begin{equation}\label{tildedi0}
\begin{aligned}
\check{d}_i(t)&\leq
\bar d_i(t)-\bar d_i(0)+\check d_i(0)+4l_{ii}^2\sum_{j=1}^N\int_0^t\tilde x_j^T(\tau)\Gamma \tilde x_j(\tau)d\tau\\
&\leq\bar d_i(t)-\bar d_i(0)+\check d_i(0)+4l_{ii}^2\sum_{j=1}^N\int_0^\infty \gamma_j\varepsilon_j(\tau)d\tau\\
&\leq\bar d_i(t)-\bar d_i(0)+\check d_i(0)+4l_{ii}^2\sum_{j=1}^N \frac{\gamma_j\varepsilon_j(0)}{k_j}.
\end{aligned}
\end{equation}
And thus
\begin{equation*}\label{etateat}
\begin{aligned}
2\tilde\xi^T(\check D\otimes\Gamma)\tilde\xi
=&\sum_{i=1}^N2\check d_i\tilde\xi_i^T\Gamma\tilde\xi_i\\
\leq&\sum_{i=1}^N\left(2\bar d_i+\beta_1\right)\tilde\xi_i^T\Gamma\tilde\xi_i,
\end{aligned}
\end{equation*}
Therefore, we can derive that
\begin{equation*}\label{xieta0}
\begin{aligned}
-2\xi^T(D\otimes\Gamma)\tilde\xi
&\leq\frac{1}{2}\xi^T(\bar D\otimes\Gamma)\xi+\frac{1}{2}\xi^T(\check D\otimes\Gamma)\xi\\
&\quad+\sum_{i=1}^N\left(4\bar d_i+\beta_1\right)\tilde\xi_i^T\Gamma\tilde\xi_i\\
&\leq\frac{1}{2}\xi^T(\bar D\otimes\Gamma)\xi+\frac{1}{2}\xi^T(\check D\otimes\Gamma)\xi\\
&\quad+\sum_{i=1}^N\left(8\bar d_i+2\beta_1\right)l_{ii}^2\sum_{j=1}^N\tilde x_j^T\Gamma \tilde x_j.
\end{aligned}
\end{equation*}
This completes the proof.
$\hfill $$\blacksquare$
\end{pf}

\begin{lem}\label{thm1-lemma2}
It holds that
\begin{equation}
\begin{aligned}
-\frac{3}{2}\xi^T(\check D\otimes\Gamma)\xi
&\leq \frac{1}{2}\xi^T(\bar D\otimes\Gamma)\xi+\xi^T(\beta_2I_N\otimes\Gamma)\xi,
\end{aligned}
\end{equation}
where $\beta_2\geq \max\{-\frac{3}{2}(\check d_i(0)+\frac{1}{3}\bar d_i(0))+6\sum_{i=1}^Nl_{ii}^2\sum_{j=1}^N\frac{\gamma_i\varepsilon_j(0)}{k_j}\}$ is a positive constant.
\end{lem}

\begin{pf}
Obviously, we have
\begin{equation*}\label{tilded}
\begin{aligned}
&-\check{d}_i(t)=-\check d_i(0)+\int_{0}^t(\xi_i^T(\tau)\Gamma\xi_i(\tau)-\hat\xi_i^T(\tau)\Gamma\hat\xi_i(\tau))d\tau\\
=&-\check d_i(0)+\int_{0}^t(-\tilde\xi_i^T(\tau)\Gamma\tilde\xi_i(\tau)-2\xi_i^T(\tau)\Gamma\tilde\xi_i(\tau))d\tau\\
\leq&-\check d_i(0)+\frac{1}{3}\int_0^t\xi_i^T(\tau)\Gamma\xi_i(\tau)d\tau
+2\int_0^t\tilde\xi_i^T(\tau)\Gamma\tilde\xi_i(\tau)d\tau\\
\leq&-\check d_i(0)+\frac{1}{3}\bar d_i(t)-\frac{1}{3}\bar d_i(0)+4l_{ii}^2\sum_{j=1}^N \frac{\gamma_j\varepsilon_j(0)}{k_j}.
\end{aligned}
\end{equation*}
Thus, we can obtain that
\begin{equation*}\label{tildedi}
\begin{aligned}
-\frac{3}{2}\xi^T(\check D\otimes\Gamma)\xi
&=\sum_{i=1}^N(-\frac{3}{2}\check d_i)\xi_i^T\Gamma\xi_i\\
&\leq\sum_{i=1}^N\Big(\frac{1}{2}\bar d_i(t)-\frac{3}{2}(\check d_i(0)+\frac{1}{3}\bar d_i(0))\\
&\quad+6\sum_{i=1}^Nl_{ii}^2\sum_{j=1}^N\frac{\gamma_i\varepsilon_j(0)}{k_j}\Big)\xi_i^T\Gamma\xi_i\\
&\leq \frac{1}{2}\xi^T(\bar D\otimes\Gamma)\xi+\xi^T(\beta_2I_N\otimes\Gamma)\xi.
\end{aligned}
\end{equation*}
This completes the proof.
$\hfill $$\blacksquare$
\end{pf}

\begin{lem}\label{thm1-lemma3}
It holds that
\begin{equation}
\begin{aligned}
-2\xi^T(\tilde D\otimes \Gamma)\hat \xi\leq 3\theta_1\sum_{i=1}^N\xi_i^T\Gamma \xi_i+4\theta_1\sum_{i=1}^Nl_{ii}^2\sum_{j=1}^N\tilde x_j^T\Gamma \tilde x_j.
\end{aligned}
\end{equation}
\end{lem}

\begin{pf}
According to the triggering rule, we have
\begin{equation*}\label{chuli_new}
\begin{aligned}
&-2\xi^T(\tilde D\otimes \Gamma)\hat \xi=-2\sum_{i=1}^N\tilde d_i\xi^T\Gamma \hat \xi_i\\
&\leq \theta_1 \sum_{i=1}^N\xi_i^T\Gamma \xi_i+\theta_1\sum_{i=1}^N\hat \xi_i^T\Gamma \hat \xi_i\\
&\leq 3\theta_1\sum_{i=1}^N\xi_i^T\Gamma \xi_i+2\theta_1\sum_{i=1}^N\tilde \xi_i^T\Gamma \tilde\xi_i\\
&\leq 3\theta_1\sum_{i=1}^N\xi_i^T\Gamma \xi_i+4\theta_1\sum_{i=1}^Nl_{ii}^2\sum_{j=1}^N\tilde x_j^T\Gamma \tilde x_j.
\end{aligned}
\end{equation*}
This completes the proof.
$\hfill $$\blacksquare$
\end{pf}

\section*{Appendix B. The lemmas in Section \ref{s4}}
For the network dynamics \dref{xi2}, we can derive the following lemmas.
\begin{lem}\label{thm5-lemma1}
It holds that
\begin{equation}\label{s3-chuli1}
\begin{aligned}
-2\mu&\xi^T[(\bar D+\bar \rho)R\mathcal L(\check D+\check \rho)\otimes \Gamma]\xi\\
&\leq \frac{\mu\lambda_2(\hat {\mathcal L})}{2N}\xi^T[(\bar D+\bar\rho)^2\otimes\Gamma]\xi\\
&+2\mu\beta_1\sigma_{\max}(R\mathcal L)\xi^T[(\bar D+\bar\rho)\otimes\Gamma]\xi\\
&+\frac{\mu\beta_1^2\sigma_{\max}(R\mathcal L)}{\alpha_1}\xi^T(I_N\otimes\Gamma)\xi,
\end{aligned}
\end{equation}
where $\alpha_1=\frac{\lambda_2(\hat L)}{4N\sigma_{\max}(R\mathcal L)}$ and $\beta_1=\check d_i(0)-\alpha_1\bar d_i(0)+2(1+\frac{1}{\alpha_1})\max\{l_{ii}^2\}\sum\limits_{j=1}^N
\left(\lambda_{\max}(Q)
+\frac{\lambda_{\max}(\Gamma)}{k_j}\right)\gamma_j\varepsilon_j(0)$.
\end{lem}

\begin{pf}
Similar to \dref{eta}, we have
\begin{equation*}\label{eta0}
\begin{aligned}
\tilde\xi_i^T\tilde\xi_i
\leq&2l_{ii}^2\sum_{j=1}^N\tilde x_j^T \tilde x_j\leq 2l_{ii}^2\sum_{j=1}^N\gamma_j\varepsilon_j.
\end{aligned}
\end{equation*}
By using the Young's inequality, we can obtain that
\begin{equation*}
\begin{aligned}
|\check{\rho}_i|=&|\hat\xi_i^TQ\hat\xi_i-\xi_i^TQ\xi_i|
=|\tilde\xi_i^TQ\tilde\xi_i+2\tilde\xi_i^TQ\xi_i|\\
\leq&\alpha_1\xi_i^TQ\xi_i+(1+\frac{1}{\alpha_1})\tilde\xi_i^TQ\tilde\xi_i\\
\leq&\alpha_1\bar\rho_i+\lambda_{\max}(Q)(1+\frac{1}{\alpha_1})\tilde\xi_i^T\tilde\xi_i\\
\leq&\alpha_1\bar\rho_i+2\lambda_{\max}(Q)(1+\frac{1}{\alpha_1})l_{ii}^2\sum_{j=1}^N\gamma_j\varepsilon_j(0),
\end{aligned}
\end{equation*}
and
\begin{equation*}
\begin{aligned}
|\check{d}_i|&=
\left|\check d_i(0)+\int_0^t(\tilde\xi_i^T\Gamma\tilde\xi_i+2\tilde\xi_i^T\Gamma\xi_i)d\tau\right|\\
&\leq\check d_i(0)+\int_0^t(\alpha_1\xi_i^T\Gamma\xi_i
+(1+\frac{1}{\alpha_1})\tilde\xi_i^T\Gamma\tilde\xi_i)d\tau\\
&\leq\check d_i(0)-\alpha_1\bar d_i(0)+\alpha_1\bar d_i\\
&\quad+\lambda_{\max}(\Gamma)(1+\frac{1}{\alpha_1})
\int_0^t\tilde\xi_i^T\tilde\xi_id\tau\\
&\leq\check d_i(0)-\alpha_1\bar d_i(0)+\alpha_1\bar d_i\\
&\quad+2\lambda_{\max}(\Gamma)(1+\frac{1}{\alpha_1})l_{ii}^2\sum_{j=1}^N \frac{\gamma_j\varepsilon_j(0)}{k_j}.
\end{aligned}
\end{equation*}
Then we have
$$
|\check{d}_i+\check{\rho}_i|\leq\alpha_1(\bar d_i+\bar\rho_i)+\beta_1.
$$
It follows that
\begin{equation}\label{37}
\begin{aligned}
&\quad-2\xi^T[(\bar D+\bar\rho)R\mathcal L(\check D+\check\rho)\otimes\Gamma]\xi\\
&\leq\frac{\lambda_2(\hat L)}{4N}\xi^T[(\bar D+\bar\rho)^2\otimes\Gamma]\xi\\
&\quad+\frac{\sigma_{\max}(R\mathcal L)}{\alpha_1}\xi^T[(\check D+\check\rho)^2\otimes\Gamma]\xi,
\end{aligned}
\end{equation}
where
\begin{equation}\label{38}
\begin{aligned}
&\frac{\sigma_{\max}(R\mathcal L)}{\alpha_1}\xi^T[(\check D+\check\rho)^2\otimes\Gamma]\xi\\
=&\frac{\sigma_{\max}(R\mathcal L)}{\alpha_1}
\sum_{i=1}^N(\check d_i+\check\rho_i)^2\xi_i^T\Gamma\xi_i\\
\leq&\frac{\sigma_{\max}(R\mathcal L)}{\alpha_1}
\sum_{i=1}^N[\alpha_1(\bar d_i+\bar\rho_i)+\beta_1]^2\xi_i^T\Gamma\xi_i\\
=&\frac{\lambda_2(\hat {\mathcal L})}{4N}\xi^T[(\bar D+\bar\rho)^2\otimes\Gamma]\xi\\
&+2\beta_1\sigma_{\max}(R\mathcal L)\xi^T[(\bar D+\bar\rho)\otimes\Gamma]\xi\\
&+\frac{\beta_1^2\sigma_{\max}(R\mathcal L)}{\alpha_1}\xi^T(I_N\otimes\Gamma)\xi.
\end{aligned}
\end{equation}
Substituting \dref{38} into \dref{37} directly yields \dref{s3-chuli1}.
$\hfill $$\blacksquare$
\end{pf}

\begin{lem}\label{thm5-lemma2}
There exists a positive time instant $T_0$ such that for $t>T_0$,
\begin{equation}
\begin{aligned}
&-2\mu\xi^T[(\bar D+\bar\rho)R\mathcal L( D+\rho)\otimes\Gamma]\tilde\xi\\
&\leq\frac{\mu\lambda_2(\hat {\mathcal L})}{4N}\xi^T[(\bar D+\bar\rho)^2\otimes\Gamma]\xi+\frac{1}{2}\mu\xi^T[(\bar D+\bar \rho)R\otimes X]\xi\\
&+\frac{\mu\sigma_{\max}(R\mathcal L)}{\alpha_1}\sum_{i=1}^N(1+\alpha_1+\beta_2)^2\bar d_i^2\tilde\xi_i^T\Gamma\tilde\xi_i,
\end{aligned}
\end{equation}
where  $X=-(QA+A^TQ-\Gamma)>0$ and $\beta_2=d_i(0)-(1+\alpha_1)\bar d_i(0)+2(1+\frac{1}{\alpha_1})\max\{l_{ii}^2\}\sum\limits_{j=1}^N
\left(\lambda_{\max}(Q)
+\frac{\lambda_{\max}(\Gamma)}{k_j}\right)\gamma_j\varepsilon_j(0)$.
\end{lem}

\begin{pf}
Obviously, we have
\begin{equation*}
\begin{aligned}
{\rho}_i=&\hat\xi_i^TQ\hat\xi_i
=\tilde\xi_i^TQ\tilde\xi_i+2\tilde\xi_i^TQ\xi_i+\xi_i^TQ\xi_i\\
\leq&(1+\alpha_1)\xi_i^TQ\xi_i+(1+\frac{1}{\alpha_1})\tilde\xi_i^TQ\tilde\xi_i\\
\leq&(1+\alpha_1)\bar\rho_i+2\lambda_{\max}(Q)(1+\frac{1}{\alpha_1})l_{ii}^2\sum_{j=1}^N\gamma_j\varepsilon_j(0),
\end{aligned}
\end{equation*}
and
\begin{equation*}
\begin{aligned}
d_i
=&d_i(0)+\int_0^t(\tilde\xi_i^T\Gamma\tilde\xi_i+2\tilde\xi_i^T\Gamma\xi_i+\xi_i^T\Gamma\xi_i)d\tau\\
\leq&d_i(0)-(1+\alpha_1)\bar d_i(0)+(1+\alpha_1)\bar d_i\\
&+(1+\frac{1}{\alpha_1})\int_0^t\tilde\xi_i^T\Gamma\tilde\xi_id\tau\\
\leq&d_i(0)-(1+\alpha_1)\bar d_i(0)+(1+\alpha_1)\bar d_i\\
&+2\lambda_{\max}(\Gamma)(1+\frac{1}{\alpha_1})l_{ii}^2\sum_{j=1}^N \frac{\gamma_j\varepsilon_j(0)}{k_j}.
\end{aligned}
\end{equation*}
Then, we have
\begin{equation*}
\begin{aligned}
d_i+{\rho}_i
\leq&(1+\alpha_1)(\bar d_i+\bar \rho_i)+\beta_2.
\end{aligned}
\end{equation*}
Thus we have
\begin{equation*}
\begin{aligned}
&-2\xi^T[(\bar D+\bar\rho)R\mathcal L( D+\rho)\otimes\Gamma]\tilde\xi\\
\leq&\frac{\lambda_2(\hat {\mathcal L})}{4N}\xi^T[(\bar D+\bar\rho)^2\otimes\Gamma]\xi\\
&~~~~~+\frac{\sigma_{\max}(R\mathcal L)}{\alpha_1}\tilde\xi^T[(D+\rho)^2\otimes\Gamma]\tilde\xi,
\end{aligned}
\end{equation*}
where
\begin{equation*}
\begin{aligned}
&\frac{\sigma_{\max}(R\mathcal L)}{\alpha_1}\tilde\xi^T[(D+\rho)^2\otimes\Gamma]\tilde\xi\\
=&\frac{\sigma_{\max}(R\mathcal L)}{\alpha_1}\sum_{i=1}^N( d_i+\rho_i)^2\tilde\xi_i^T\Gamma\tilde\xi_i\\
\leq&\frac{\sigma_{\max}(R\mathcal L)}{\alpha_1}
\sum_{i=1}^N[(1+\alpha_1)(\bar d_i+\bar\rho_i)+\beta_2]^2\tilde\xi_i^T\Gamma\tilde\xi_i\\
=&\frac{\sigma_{\max}(R\mathcal L)}{\alpha_1}\sum_{i=1}^N[(1+\alpha_1)^2\bar d_i^2+\\
&~~~~~~~~~~~~~~~~~~2(1+\alpha_1)\beta_2\bar d_i+\beta_2^2]\tilde\xi_i^T\Gamma\tilde\xi_i\\
&+\frac{\sigma_{\max}(R\mathcal L)}{\alpha_1}\sum_{i=1}^N[
(1+\alpha_1)^2\bar\rho_i+\\
&~~~~~~~~~~~~~~~~~~2(1+\alpha_1)^2\bar d_i+2\beta_2(1+\alpha_1)]\bar\rho_i\tilde{\xi}_i^T\Gamma\tilde\xi_i.
\end{aligned}
\end{equation*}
Since $\bar d_i(t)\geq1$, we can derive that
\begin{equation*}\label{ineqe}
\begin{aligned}
&[(1+\alpha_1)^2\bar d_i^2+2(1+\alpha_1)\beta_2\bar d_i+\beta_2^2]
\tilde\xi_i^T\Gamma\tilde\xi_i\\
\leq&[(1+\alpha_1)^2+2(1+\alpha_1)\beta_2+\beta_2^2]\bar d_i^2
\tilde\xi_i^T\Gamma\tilde\xi_i\\
\leq&2(1+\alpha_1+\beta_2)^2\lambda_{\max}(\Gamma)l_{ii}^2\bar d_i^2\sum_{j=1}^N\tilde x_j^T \tilde x_j,
\end{aligned}
\end{equation*}
and
\begin{equation*}\label{ineqxi}
\begin{aligned}
&[(1+\alpha_1)^2\bar\rho_i+
2(1+\alpha_1)^2\bar d_i+2\beta_2(1+\alpha_1)]\bar\rho_i\tilde{\xi}_i^T\Gamma\tilde\xi_i\\
\leq&\max\{(1+\alpha_1)^2,2(1+\alpha_1)(1+\alpha_1+\beta_2)\}(\bar d_i+\bar\rho_i)\bar\rho_i
\tilde\xi_i^T\Gamma\tilde\xi_i\\
\leq&\alpha_2(\bar d_i+\bar\rho_i)\bar \rho_i\sum_{j=1}^N\gamma_j\varepsilon_j(t),
\end{aligned}
\end{equation*}
where $\alpha_2=2\max\{(1+\alpha_1)^2,2(1+\alpha_1)(1+\alpha_1+\beta_2)\}
\lambda_{\max}(\Gamma)l_{ii}^2$.
Since $\varepsilon_j(t)\leq\exp(-k_jt)\varepsilon_j(0)$, there exists a time instant $T_0$ such that
$$
\alpha_2\sum_{j=1}^N\gamma_j\varepsilon_j(t)
\leq\frac{\alpha_1\lambda_{\min}(R)\lambda_{\min}(X)}{2\sigma_{\max}(R\mathcal L)\lambda_{\max}(Q)},
$$
holds for all $t>T_0$. Then we obtain that for $t>T_0$,
\begin{equation*}
\begin{aligned}
&\frac{\sigma_{\max}(RL)}{\alpha_1}\sum_{i=1}^N[
(1+\alpha_1)^2\bar\rho_i+2(1+\alpha_1)^2\bar d_i+2\beta_2(1+\alpha_1)]\\&\quad\bar\rho_i\tilde{\xi}_i^T\Gamma\tilde\xi_i
\leq\frac{1}{2}\xi^T[(\bar D+\bar\rho)R\otimes X]\xi.
\end{aligned}
\end{equation*}
Thus, for $t>T_0$, we have
\begin{equation*}\label{s3-chuli2}
\begin{aligned}
&-2\mu\xi^T[(\bar D+\bar\rho)R\mathcal L( D+\rho)\otimes\Gamma]\tilde\xi\\
\leq&\frac{\mu\lambda_2(\hat {\mathcal L})}{4N}\xi^T[(\bar D+\bar\rho)^2\otimes\Gamma]\xi+\frac{1}{2}\mu\xi^T[(\bar D+\bar \rho)R\otimes X]\xi\\
&\quad+\frac{\mu\sigma_{\max}(R\mathcal L)}{\alpha_1}\sum_{i=1}^N(1+\alpha_1+\beta_2)^2\bar d_i^2\tilde\xi_i^T\Gamma\tilde\xi_i.
\end{aligned}
\end{equation*}
This completes the proof.
$\hfill $$\blacksquare$
\end{pf}

\begin{lem}\label{thm5-lemma3}
It holds that
\begin{equation}
\begin{aligned}
&-2\mu\xi^T[(\bar D+\bar \rho)R\mathcal L\tilde D\otimes \Gamma]\hat \xi\\
&\leq \frac{\mu\lambda_2(\hat {\mathcal L})}{8N}\sigma_{\max}(R\mathcal L)\xi^T[(\bar D+\bar \rho)^2\otimes \Gamma]\xi\\
&+\frac{16N\mu\theta^2}{\lambda_2(\hat {\mathcal L})}\sigma_{\max}(R\mathcal L)\xi^T(I_N\otimes \Gamma)\xi\\
&+\frac{16N\mu\theta^2}{\lambda_2(\hat {\mathcal L})}\sigma_{\max}(R\mathcal L)\tilde\xi^T(I_N\otimes \Gamma)\tilde\xi.
\end{aligned}
\end{equation}
\end{lem}

\begin{pf}
According to the triggering rule, we can obtain that
\begin{equation*}\label{s3-chuli3}
\begin{aligned}
&-2\mu\xi^T[(\bar D+\bar \rho)R\mathcal L\tilde D\otimes \Gamma]\hat \xi\\
&\leq \frac{\mu\lambda_2(\hat {\mathcal L})}{8N}\sigma_{\max}(R\mathcal L)\xi^T[(\bar D+\bar \rho)^2\otimes \Gamma]\xi\\
&+\frac{8N\mu}{\lambda_2(\hat {\mathcal L})}\sigma_{\max}(R\mathcal L)\hat\xi^T[\tilde D^2\otimes \Gamma]\hat \xi\\
&\leq \frac{\mu\lambda_2(\hat {\mathcal L})}{8N}\sigma_{\max}(R\mathcal L)\xi^T[(\bar D+\bar \rho)^2\otimes \Gamma]\xi\\
&+\frac{16N\mu\theta^2}{\lambda_2(\hat {\mathcal L})}\sigma_{\max}(R\mathcal L)\xi^T(I_N\otimes \Gamma)\xi\\
&+\frac{16N\mu\theta^2}{\lambda_2(\hat {\mathcal L})}\sigma_{\max}(R\mathcal L)\tilde\xi^T(I_N\otimes \Gamma)\tilde\xi.
\end{aligned}
\end{equation*}
This completes the proof.
$\hfill $$\blacksquare$
\end{pf}

\bibliographystyle{plain}
\bibliography{IFACbibfile}

\begin{thebibliography}{10}

\bibitem{Alm2017TAC}
J.~Almeida, C.~Silvestre, and A.~Pascoal.
\newblock Synchronization of multiagent systems using event-triggered and
  self-triggered broadcasts.
\newblock {\em IEEE Trans. Autom. Control}, 62(9):4741--4746, 2017.

\bibitem{DBernstein2009Matrix}
D.~Bernstein.
\newblock {\em Matrix Mathematics: Theory, Facts, and Formulas}.
\newblock Princeton University Press, Princeton, NJ, 2009.

\bibitem{BCheng2019tac}
B.~Cheng and Z.~Li.
\newblock Coordinated tracking control with asynchronous edge-based
  event-triggered communications.
\newblock {\em IEEE Trans. Autom. Control}, 64(10):4321--4328, 2019.

\bibitem{BCheng2018fully}
B.~Cheng and Z.~Li.
\newblock Fully distributed event-triggered protocols for linear multi-agent
  networks.
\newblock {\em IEEE Trans. Autom. Control}, 64(4):1655--1662, 2019.

\bibitem{BCheng2020CDC}
B.~Cheng, Y.~Lv, and Z.~Li.
\newblock Distributed adaptive event-triggered consensus with discrete control
  updating.
\newblock {\em 59th IEEE Conf. Decis. Control}, pages 2793--2798, 2020.

\bibitem{thcheng2017event}
T.~Cheng, Z.~Kan, J.~R. Klotz, J.M. Shea, and W.E. Dixon.
\newblock Event-triggered control of multiagent systems for fixed and
  time-varying network topologies.
\newblock {\em Automatica}, 62(10):5365--5371, 2017.

\bibitem{Cheng2016Event}
Y.~Cheng and V.~Ugrinovskii.
\newblock Event-triggered leader-following tracking control for multivariable
  multi-agent systems.
\newblock {\em Automatica}, 70:204--210, 2016.

\bibitem{DVDimarogonas2012distributed}
D.V. Dimarogonas, E.Frazzoli, and K.~Johansson.
\newblock Distributed event-triggered control for multi-agent systems.
\newblock {\em IEEE Trans. Autom. Control}, 57(5):1291--1297, 2012.

\bibitem{Dolk2017TAC}
V.~Dolk, D.P. Borgers, and W.P.M.H. Heemels.
\newblock Output-based and decentralized dynamic event-triggered control with
  guaranteed $l_p$-gain performance and zeno-freeness.
\newblock {\em IEEE Trans. Autom. Control}, 62(1):34--49, 2017.

\bibitem{Garcia2013decentralised}
E.~Garcia, Y.~Cao, H.~Yu, A.~Giua, P.~Antsaklis, and D.~Casbeer.
\newblock Decentralised event-triggered cooperative control with limited
  communication.
\newblock {\em Int. J. Control}, 86(9):1479--1488, 2013.

\bibitem{Girard2015TAC}
A.~Girard.
\newblock Dynamic triggering mechanisms for event-triggered control.
\newblock {\em IEEE Trans. Autom. Control}, 60(7):1992--1997, 2015.

\bibitem{Heemels2013TAC}
W.P.M.H. Heemels, M.C.F. Donkers, and A.R. Teel.
\newblock Periodic event-triggered control for linear systems.
\newblock {\em IEEE Trans. Autom. Control}, 58(4):847--861, 2013.

\bibitem{WHu2017output}
W.~Hu, L.~Liu, and G.~Feng.
\newblock Output consensus of heterogeneous linear multi-agent systems by
  distributed event-triggered/self-triggered strategy.
\newblock {\em IEEE Trans. Cybern.}, 47(8):1914--1924, 2017.

\bibitem{NaHuang2020IS}
N.~Huang, Z.~Sun, B.D.D. Anderson, and Z.~Duan.
\newblock Distributed and adaptive triggering control for networked agents with
  linear dynamics.
\newblock {\em Information Science}, 517:297--314, 2020.

\bibitem{PIoannou1996robust}
P.~Ioannou and J.~Sun.
\newblock {\em Robust Adaptive Control}.
\newblock Prentice-Hall, New York, NY, 1996.

\bibitem{XianweiLi2020}
X.~Li, Z.~Sun, Y.~Tang, and H.R. Karimi.
\newblock Adaptive event-triggered consensus of multi-agent systems on directed
  graphs.
\newblock {\em IEEE Trans. Autom. Control}, in press, doi:
  10.1109/TAC.2020.3000819, 2020.

\bibitem{XianweiLi2020Automatica}
X.~Li, Y.~Tang, and H.R. Karimi.
\newblock Consensus of multi-agent systems via fully distributed
  event-triggered control.
\newblock {\em Automatica}, 116:108898, 2020.

\bibitem{ZLi2014cooperative}
Z.~Li and Z.~Duan.
\newblock {\em Cooperative Control of Multi-agent Systems: A Consensus Region
  Approach}.
\newblock CRC Press, Boca Raton, FL, 2014.

\bibitem{Liu2017distributed}
T.~Liu, M.~Cao, C.D. Persis, and J.M. Hendrickx.
\newblock Distributed event-triggered control for asymptotic synchronization of
  dynamical networks.
\newblock {\em Automatica}, 86:199--204, 2017.

\bibitem{YuezuLv2016Automatica}
Y.~Lv, Z.~Li, Z.~Duan, and J.~Chen.
\newblock Distributed adaptive output feedback consensus protocols for linear
  systems on directed graphs with a leader of bounded input.
\newblock {\em Automatica}, 74:308--314, 2016.

\bibitem{YuezuLv2020Automatica}
Y.~Lv, G.~Wen, T.~Huang, and Z.~Duan.
\newblock Adaptive attack-free protocol for consensus tracking with pure
  relative output information.
\newblock {\em Automatica}, 117:108998, 2020.

\bibitem{JieMei2016TAC}
J.~Mei, W.~Ren, and J.~Chen.
\newblock Distributed consensus of second-order multi-agent systems with
  heterogeneous unknown inertias and control gains under a directed graph.
\newblock {\em IEEE Trans. Autom. Control}, 61(8):2019--2034, 2016.

\bibitem{Meng2013event}
X.~Meng and T.~Chen.
\newblock Event based agreement protocols for multi-agent networks.
\newblock {\em Automatica}, 49(7):2125--2132, 2013.

\bibitem{Nowzari2019Auto}
C.~Nowzari, E.~Garcia, and J.~Cort\'{e}s.
\newblock Event-triggered communication and control of networked systems for
  multi-agent consensus.
\newblock {\em Automatica}, 105:1--27, 2019.

\bibitem{YangyangQian2019TAC}
Y.~Qian, L.~Liu, and G.~Feng.
\newblock Output consensus of heterogeneous linear multi-agent systems with
  adaptive event-triggered control.
\newblock {\em IEEE Trans. Autom. Control}, 64(6):2606--2613, 2019.

\bibitem{Seyboth2013event}
G.~Seyboth, D.~Dimarogonas, and K.~Johanasson.
\newblock Event-based broadcasting for multi-agent average consensus.
\newblock {\em Automatica}, 49(1):245--252, 2013.

\bibitem{LXing2017event}
L.~Xing, C.~Wen, F.~Guo, Z.~Liu, and H.~Su.
\newblock Event-based consensus for linear multiagent systems without
  continuous communication.
\newblock {\em IEEE Trans. Cybern.}, 47(8):2132--2142, 2017.

\bibitem{DYang2016Decentralized}
D.~Yang, W.~Ren, X.~Liu, and W.~Chen.
\newblock Decentralized event-triggered consensus for linear multi-agent
  systems under general directed graphs.
\newblock {\em Automatica}, 69:242--249, 2016.

\bibitem{XYi2019TAC}
X.~Yi, K.~Liu, D.V. Dimarogonas, and K.H. Johansson.
\newblock Dynamic event-triggered and self-triggered control for multi-agent
  systems.
\newblock {\em IEEE Trans. Autom. Control}, 64(8):3300--3307, 2019.

\bibitem{HYu2020TAC}
H.~Yu and T.~Chen.
\newblock On zeno behavior in event-triggered finite-time consensus of
  multi-agent systems.
\newblock {\em IEEE Trans. Autom. Control}, in press, doi:
  10.1109/TAC.2020.3030758, 2020.

\bibitem{Yu2016leader}
P.~Yu, L.~Ding, Z.~Liu, and Z.~Guan.
\newblock Leader-follower flocking based on distributed event-triggered hybrid
  control.
\newblock {\em Int. J. Robust Nonlinear Control}, 26:143--153, 2016.

\bibitem{PYu2018TAC}
P.~Yu, C.~Fischione, and D.V. Dimarogonas.
\newblock Distributed event-triggered communication and control of linear
  multiagent systems under tactile communication.
\newblock {\em IEEE Trans. Autom. Control}, 63(11):3979--3985, 2018.

\bibitem{HZhang2014observer}
H.~Zhang, G.~Feng, H.~Yan, and Q.~Chen.
\newblock Observer-based output feedback event-triggered control for consensus
  of multi-agent systems.
\newblock {\em IEEE Trans. Ind. Electron.}, 61(9):4885--4894, 2014.

\bibitem{Zhu2014event}
W.~Zhu, Z.~Jiang, and G.~Feng.
\newblock Event-based consensus of multi-agent systems with general linear
  models.
\newblock {\em Automatica}, 50(2):552--558, 2014.

\end{thebibliography}

\end{document}